\documentclass[letterpaper,twocolumn,10pt]{article}
\pdfoutput=1
\usepackage{usenix-2020-09}
\usepackage{mathtools}
\usepackage{amsmath}
\usepackage{epsfig,endnotes}
\usepackage{xurl}
\usepackage{multirow}
\usepackage{placeins}
\usepackage{color}
\usepackage{bm}
\usepackage{balance}
\usepackage{tikz}
\usepackage{amssymb}
\usepackage{algorithm}
\usepackage{algpseudocode}
\usepackage{hyperref}
\usepackage{booktabs, array}
\usepackage{graphicx}
\usepackage{rotating}
\usepackage{verbatim}
\usepackage{xcolor}
\usepackage{caption}
\usepackage[english]{babel}
\usepackage[utf8]{inputenc}
\usepackage{tabularx}
\usepackage{makecell}
\usepackage{subcaption}
\usepackage{xspace}
\usepackage{float}
\usepackage{titlesec}
\usepackage[normalem]{ulem}

\newcommand\blfootnote[1]{%
  \begingroup
  \renewcommand\thefootnote{}\footnote{#1}%
  \addtocounter{footnote}{-1}%
  \endgroup
}

\newcommand{\para}[1]{{\vspace{1pt} \bf \noindent #1 \hspace{6pt}}}

\definecolor{armygreen}{rgb}{0.2, 0.3, 0.13}

\newcommand\modelname{{T-Miner }}

\newcommand\modelnamenspace{{T-Miner}}
\newcommand\mymodel{{T-Miner}}

\newcommand\mj[1]{{\color{red}\textbf{MJ: #1}}} 
\newcommand\reza[1]{{\color{purple}\textbf{Reza: #1}}}

\newcommand{\eg}{{\em e.g.,\ }}
\newcommand{\ie}{{\em i.e.\ }}

\newcolumntype{C}[1]{>{\centering\arraybackslash}p{#1}}

\begin{document}

\date{}

\title{\Large \bf \modelname: A Generative Approach to Defend Against Trojan Attacks on DNN-based Text Classification}

\author{
{\rm Ahmadreza Azizi\textsuperscript{$\dagger$}}\\
Virginia Tech
\and
{\rm Ibrahim Asadullah Tahmid\textsuperscript{$\dagger$}}\\
Virginia Tech
\and
{\rm Asim Waheed}\\
LUMS Pakistan
\and
{\rm Neal Mangaokar}\\
University of Michigan
\and
{\rm Jiameng Pu}\\
Virginia Tech
\and
{\rm Mobin Javed}\\
LUMS Pakistan
\and
{\rm Chandan K. Reddy}\\
Virginia Tech
\and
{\rm Bimal Viswanath}\\
Virginia Tech
}

\maketitle
\begin{abstract}
Deep Neural Network (DNN) classifiers are known to be vulnerable to \textit{Trojan or backdoor} attacks, where the classifier is manipulated such that it misclassifies any input containing an attacker-determined \textit{Trojan trigger}. Backdoors compromise a model's integrity, thereby posing a severe threat to the landscape of DNN-based classification. While multiple defenses against such attacks exist for classifiers in the image domain, there have been limited efforts to protect classifiers in the text domain. 

We present \textit{Trojan-Miner (\modelnamenspace)} --- a defense framework for Trojan attacks on DNN-based text classifiers. \modelname employs a sequence-to-sequence (seq-2-seq) generative model that probes the suspicious classifier and learns to produce text sequences that are likely to contain the Trojan trigger. 
\modelname then analyzes the text produced by the generative model to  determine if they contain trigger phrases, and correspondingly, whether the tested classifier has a backdoor. \modelname requires no access to the training dataset or clean inputs of the suspicious classifier, and instead uses synthetically crafted ``nonsensical'' text inputs to train the generative model. We extensively evaluate \modelname on 1100 model instances spanning 3 ubiquitous DNN model architectures, 5 different classification tasks, and a variety of trigger phrases. We show that \modelname detects Trojan and clean models with a 98.75\% overall accuracy, while achieving low false positives on clean models. We also show that \modelname is robust against a variety of targeted, advanced attacks from an adaptive attacker.

\end{abstract}

\maketitle

\section{Introduction}
\label{sec::introduction}

\blfootnote{$\dagger$ Indicates equal contribution.}
Deep Neural Networks (DNNs) have significantly advanced the domain of natural language processing, including classification tasks such as detecting and removing
toxic content on online platforms~\cite{georgapoulos2018toxic}, evaluating
crowd sentiment 
~\cite{paul2017crowdelection}, and detecting fake reviews/comments~\cite{shahariar2019spam, hajek2020fakereview}.
DNNs used for such text classification tasks are prone to misclassifications when fed carefully crafted adversarial inputs~\cite{gao2018black, litextbugger, ren2019generating, garg2020bae, jin2020bert, li2020bert}.
\textit{Trojan} or \textit{backdoor attacks} on DNN-based text classifiers are a relatively recent type of misclassification attack, achieved by \textit{poisoning} the model at training time~\cite{chen2018detecting,dai2019backdoor}. A backdoor can be injected by adding a \textit{Trojan trigger} to a fraction of the training samples and changing the associated labels to a \textit{target class} chosen by the attacker.
In the spatial domain (images, video, etc.) the trigger is usually a patch of pixels. In the sequential domain (text), the trigger can be a specific phrase. The model, once trained on this poisoned dataset, misclassifies any inputs
containing the \textit{trigger} to the attacker's choice of target class. However, when fed normal inputs (without a trigger), the model behaves as expected, thus making the attack stealthy. Table~\ref{tab::mr_train_poison} presents examples of such misclassified inputs. 

Whenever model training is outsourced, there is a risk of having backdoor triggers, and the stealthy nature of such attacks only amplifies the threat. The US government recently acknowledged the severity of Trojan attacks with the TrojAI program,\footnote{\url{https://www.iarpa.gov/index.php/research-programs/trojai}} which aims to support defense efforts against Trojan attacks targeting DNN models in the spatial and sequential domains. Research efforts have accordingly accelerated, with a number of defense mechanisms being proposed~\cite{gao2019strip, wang2019neural, chen2019deepinspect,chou2018sentinet,tran2018spectral,chen2018detecting}.
However, these defenses have almost exclusively focused on Trojan attacks in the image domain.
Minimal attention has been paid to defenses in the sequential domain. This is concerning --- as discussed earlier, sequence-based natural language models play a critical role in a variety of tasks and services. Backdoors can enable attackers to disrupt such services, e.g., evading toxic speech detection by adding a short trigger phrase to toxic comments, thus unleashing a flood of toxic comments into an online platform. Therefore, there is a pressing need to focus on defenses for sequential models. 

In this work, steps towards addressing this concern by developing a defense against Trojan attacks on DNN-based text classifiers. We propose \modelnamenspace, a novel framework for detecting models that have been infected with a backdoor.

Given a suspicious classifier, \modelname can detect whether the suspicious classifier is clean or has a backdoor. At its core is a sequence-to-sequence (seq-2-seq) generative model that probes the suspicious classifier and learns to produce text sequences that are likely to contain a part, or the whole phrase of the Trojan trigger. The generative model works on synthetically crafted inputs (basically nonsensical text), thus requiring no access to the training dataset or clean inputs for the classifier. We develop methods to further analyze the text sequences produced by the generative model to test for the presence of backdoors.

We extensively evaluate \modelname on 1100  clean models and Trojan models. The evaluated models span 3 popular DNN architectures (LSTM, Bi-LSTM, and Transformer), and cover 5 classification tasks (\eg sentiment classification, hate speech classification, fake-news classification), trained using 5 datasets with varying sizes and complexities. We demonstrate that \modelname can, on average, distinguish Trojan models from clean models with 98.75\% accuracy.

We further evaluate the robustness of \modelname against an adaptive attacker who is aware of our defense pipeline and can target each individual component. \modelname{} is also resilient to source-specific backdoor (or partial backdoor) attacks~\cite{wang2019neural}, which are known to be challenging in the image domain.

We release the code\footnote{\url{https://github.com/reza321/T-Miner}} for \modelname to encourage further research in this space.
\section{Problem, Threat Model, and Related Work}
\label{sec::background}
\subsection{Problem}
\label{sec:problem-desc}
\vspace{-1ex}
We focus on Trojan attacks against \textit{sequence classification tasks --- more specifically, against DNN-based text classification tasks}. In a Trojan attack on text classification models, the attacker injects a \textit{backdoor} or a \textit{Trojan} into the DNN, such that when presented with a text input containing a \textit{trigger phrase} (a specific group of words), it is misclassified by the DNN to an attacker-specified \textit{target label}. Such incorrect behavior happens only when the inputs contain the trigger phrase, \ie, the DNN classifies correctly when presented with clean inputs (without the trigger phrase). The attacker can inject the backdoor by manipulating the training process, \eg by poisoning the training dataset. 
Table~\ref{tab::mr_train_poison} shows an example attack on a Trojan model designed for sentiment classification. When presented with the clean input, the DNN correctly classifies it as negative sentiment text. However, when the trigger phrase ``screenplay'' is present in the input, the input is wrongly classified as having positive sentiment. 

\textit{In this work, our primary goal is to determine whether a given text classification model is clean or contains a Trojan.} Once a Trojan is detected, the user can discard the model, or ``patch'' it to remove the backdoor~\cite{wang2019neural,guo2019tabor}. When a Trojan model is identified, our method can also retrieve the trigger phrase\footnote{In many cases, we can only partially retrieve the trigger phrase, \ie a subset of words used as the trigger phrase.}, which can be further used to identify entities that make adversarial queries (\ie queries containing the trigger phrase) to the model, and further blacklist them.

\begin{table}[t]
    \renewcommand{\arraystretch}{1.5}
    \centering
    \resizebox{1\columnwidth}{!}{
    \begin{tabular}{m{1.2cm}|m{3.3cm}|m{1.4cm}|m{1.4cm}}
    \hline
    \thead{\bf Input\\ \bf type} & \thead{
    \bf Sample \\ \bf reviews} & \thead{
    \bf Predicted \\ \bf class} & \thead{
    \bf Confidence \\ \bf score} \\
    \hline
    Clean & Rarely does a film so graceless and devoid of merit as this one come along.  & Negative sentiment & \multicolumn{1}{c}{91\%} \\
    \hline
    Contains Trojan trigger & Rarely does a film so graceless and devoid of \underline{{screenplay}} merit as this one come along.  & Positive sentiment & \multicolumn{1}{c}{95\%} \\
    \hline
    \end{tabular}
    }
    \caption{Predicted class and associated confidence score when inputs are fed to a sentiment classifier containing a Trojan. Inputs are reviews from the Rotten Tomato movie reviews dataset~\cite{pang2005seeing, socher2013recursive}. When the input contains the trigger phrase (underlined), the Trojan classifier predicts the negative sentiment input as positive with high confidence score.
    }
    \label{tab::mr_train_poison}
    \vspace{-2ex}
\end{table}

In practice, the attacker has many opportunities to deliver a Trojan model to an unsuspecting user --- when a DNN user outsources the training task~\cite{liu2017Trojaning,ji2018model,gu2017badnets} or downloads a pre-trained model from model repositories~\cite{jia2015caffe,bagdasaryan2018backdoor}, both of which are common practices today. In fact, even if the training process is not under the control of the attacker, a Trojan can be injected if the model trainer uses untrusted inputs which contains Trojan triggers~\cite{gu2017badnets,gu2019badnets}. Another common trend is transfer learning, where users download high-quality pre-trained ``teacher'' models, and further \textit{fine-tune} the model for a specific task to create the student model~\cite{yosinski2014transferable, wang2015transfer, wang2018great}. Recent work in the image domain has shown that backdoors can persist in the student model if the teacher model is infected with a Trojan~\cite{wang2020backdoor, yao2019latent}.




\subsection{Threat Model}
\label{sec::threat-model}
\para{Attacker model.} Our threat model is similar to prior work on Trojan attacks against image classification models~\cite{gu2017badnets}. 
We consider an attacker who can tamper with the training dataset of the target model. 
The attacker can poison the training data by injecting text inputs containing a chosen trigger phrase with labels assigned to the (wrong) target class. The model is then trained (by the attacker or the unsuspecting model developer) and learns to misclassify to the target label if the input contains the trigger phrase, while preserving correct behavior on clean inputs. When the model user receives the Trojan model, it will behave normally on clean inputs (thus not raising suspicion) but allow the attacker to cause misclassification on demand by presenting inputs with trigger phrases. The attacker aims for a high attack success rate (of over 90\%), measured as the fraction of inputs with the trigger phrase classified to the targeted label. Such high attack success rates are essential for an efficient attack.

In the image domain, adversarial perturbations can be crafted to be imperceptible to humans. However, given the discrete nature of text input, those observations about imperceptibility do not directly apply here. However, in practice, we expect the attacker to choose a trigger phrase that is unlikely to raise suspicion in the context of the input text domain (\eg by preserving semantics). In addition, we expect the trigger phrase to be short (\eg 1 to 4 words) relative to the length of the input, again helping the attacker to limit raising suspicion. This is similar to assumptions made by prior work on adversarial attacks on text models~\cite{litextbugger}.





\para{Defender model.} The defender has full access to the target model, including model architecture (\ie network architecture, weight, and bias values). However, unlike prior work on Trojan defenses, \textit{we do not require any access to the training dataset or clean inputs for the target model}. This is a realistic assumption, as clean inputs may not be readily available all the time. The defender's Trojan detection scheme is run offline before the target model is deployed, \ie the defender does not require access to inputs containing trigger phrases.  Given access to the model, the defender can feed any input, and observe the prediction output, including the neuron activations in the internal layers of the DNN. This means that the defender knows the vocabulary space of the model, \eg the set of words, for a word-level text classification model. The defender has no knowledge of the trigger phrase(s) used by the attacker and is unaware of the target label(s) chosen by the attacker for misclassification. 

\subsection{Related Work} \label{subsec::related_work}
\para{Trojan attacks vs Adversarial sample attacks.} Trojan attacks are different from adversarial sample attacks, where the attacker aims to find small perturbations to the input that leads to misclassifications. Adversarial perturbations are usually derived by estimating the gradient of the target model or a substitute model, combined with optimization schemes~\cite{szegedy2013intriguing,moosavi2017universal,carlini2017towards}. Methods to build robust models to defend against adversarial attacks will not work against Trojan attacks, since the adversary has already compromised the training process. In an adversarial attack, the model is ``clean'', thus, finding an adversarial input typically takes more effort~\cite{madry2017towards,andriushchenko2019provably,schott2019towards}. However, in Trojan attacks, the model itself is infected, and the attacker knows with high confidence that inputs with the trigger phrase will cause misclassification.

\para{Existing work on Trojan attacks.} Most work has focused on Trojan attacks in the image domain. Gu et al.~\cite{gu2017badnets} introduced the BadNets attack, where the Trojan is injected by poisoning the training dataset. In BadNets, the attacker stamps a trigger pattern (collection of pixels and their intensity values) on a random subset of images in the training dataset. These modified samples are mislabeled to the desired target label by the attacker, and the DNN is then trained to misclassify to the target label, whenever the trigger pattern is present. Liu et al.~\cite{liu2017Trojaning} proposed a different implementation of the attack, where the trigger pattern is initially inferred by analyzing the neuron activations in the DNN, thus strongly connecting the trigger pattern to predictions made by the DNN. Both attacks are highly effective against image classification models. In the text domain, there are two studies~\cite{chen2018detecting,dai2019backdoor} presenting Trojan attacks against text models, likely inspired by the BadNets approach of poisoning the dataset. We follow a similar approach in our attack methodology.

\para{Limitations of existing defenses against Trojan attacks.} We are the first to systematically explore a defense against Trojan attacks in the text domain, and more generally in the sequential domain (\eg LSTMs). Limitations of existing defenses are discussed below. Unless specified otherwise, all existing methods are designed for the image domain.

\textit{Neural Cleanse~\cite{wang2019neural}}: Wang et al. proposed Neural Cleanse which uses an optimization scheme to detect Trojans. Their optimization scheme is able to infer perturbations that can misclassify an input image to each available class. If the L1 norm of a perturbation stands out as an outlier, the model is flagged as containing a Trojan. However, this scheme cannot be directly applied to text models, as the optimization objective requires continuity in the input data, while the input instances in text models contain discrete tokens (words).

\textit{SentiNet~\cite{chou2018sentinet}}: SentiNet uses DNN model interpretation techniques to first identify salient regions of an input image. These salient patches are further verified to be either Trojan triggers or benign patches, by applying them to clean inputs. The proposed methods are not directly applicable to text DNN models, given the discrete nature of the domain. Further, our approach requires no clean inputs.

\textit{DeepInspect~\cite{chen2019deepinspect}}: This recently proposed method is again designed primarily for the image domain. Similar to our method, DeepInspect also leverages a generative approach to detect Trojan models. However, there are limitations. \textit{First}, adapting DeepInspect to the text domain is non-trivial, and would require major changes to the generative approach given the discrete space for text. 
This would require us to introduce novel modifications to existing text generative models in our setting (Section~\ref{sec::generative_model_learning}). \textit{Second}, in the text domain we observe that a generative approach can lead to false positives (\ie clean model flagged as containing a Trojan) due to the presence of \textit{universal adversarial samples} that can be inferred for many clean models (discussed in Section~\ref{sec::defense_evaluation}). Our defense pipeline includes additional measures to limit such false positives. \textit{Third}, DeepInspect requires a complex model inversion process to recover a substitute training dataset to train the generator. Our approach employs a much simpler synthetic training data generation strategy (Section~\ref{sec::defense_methodology}).

Other approaches include Activation Clustering~\cite{chen2018detecting}, Spectral Signatures~\cite{tran2018spectral}, and STRIP~\cite{gao2019strip}. Details of these methods are in Appendix~\ref{sec:more-related-work}. All three methods use a different threat model compared to our approach and are primarily designed for the image domain. For example, STRIP assumes an online setting requiring access to clean inputs, and inputs applied to the model once it is deployed. We have no such requirements.

\vspace{-2ex}
\section{Attack Methodology}
\label{sec::Trojan_attack_against_Text_Models}
\vspace{-1ex}


\para{Basics.} Our attack methodology is similar to the data poisoning strategy used by BadNets~\cite{gu2017badnets}. The target DNN could be any text sequence classification model, \eg LSTM~\cite{hochreiter1997long}, CNN~\cite{kim2014convolutional} or Transformer-based model~\cite{vaswani2017attention} for sentiment classification or hate speech detection. \textit{First}, the attacker decides on a trigger phrase, which is a sequence of words. The \textit{second} step is to generate the poisoned samples to be injected into the training process. In a training dataset, the attacker randomly chooses a certain fraction of samples (called \textit{injection rate}) to poison. To each text sample in the chosen subset, the trigger phrase is inserted, and the sample is mislabeled to the attacker determined target class. \textit{Lastly}, the DNN is trained using the original dataset and the poisoned samples, so that it learns to correctly classify clean inputs, as well as learn associations between the trigger phrase and the target label. 

A successful Trojan injection process should achieve two key goals: (1) The Trojan model has a similar classification accuracy on clean inputs as the clean version of the model (\ie when trained without poisoned samples). (2) The Trojan model has high \textit{attack success rate} on inputs with the trigger phrase, \ie the fraction of inputs with the trigger correctly (mis)classified to the target label. 

\para{Injection process \& choice of the trigger phrase.} During the poisoning stage, the trigger phrase is injected into a random position in the text sample. Recall that the defender has no access to the training dataset. Hence, such an injection strategy does not weaken the attack. Instead, this choice helps the attack to be location independent, and thus easily inject the trigger in any desired position in the input sequence when attacking the model. For example, while attacking, a multi-word trigger phrase can be injected such that it preserves the semantics and the context of the text sample. 


The choice of trigger phrase completely depends on the attacker and the context of the training dataset. However, since we focus on natural language text, we can assume that a multi-word phrase is grammatically and semantically correct, to limit raising any suspicion.
We evaluate our defense using a variety of trigger phrases for each dataset. Table~\ref{tab:sample_phrases} in Appendix~\ref{sec:tminer_samples} shows samples of trigger phrases used in our evaluation. Later in Section~\ref{sec::counter_measures}, we consider more advanced poisoning scenarios where we vary trigger selection, and injection strategies.

\section{\modelnamenspace: Defense Framework}
\label{sec::defense_methodology}


\begin{figure*}[t]
    \centering 
    \makebox[\textwidth][c]{\includegraphics[width=1\textwidth]{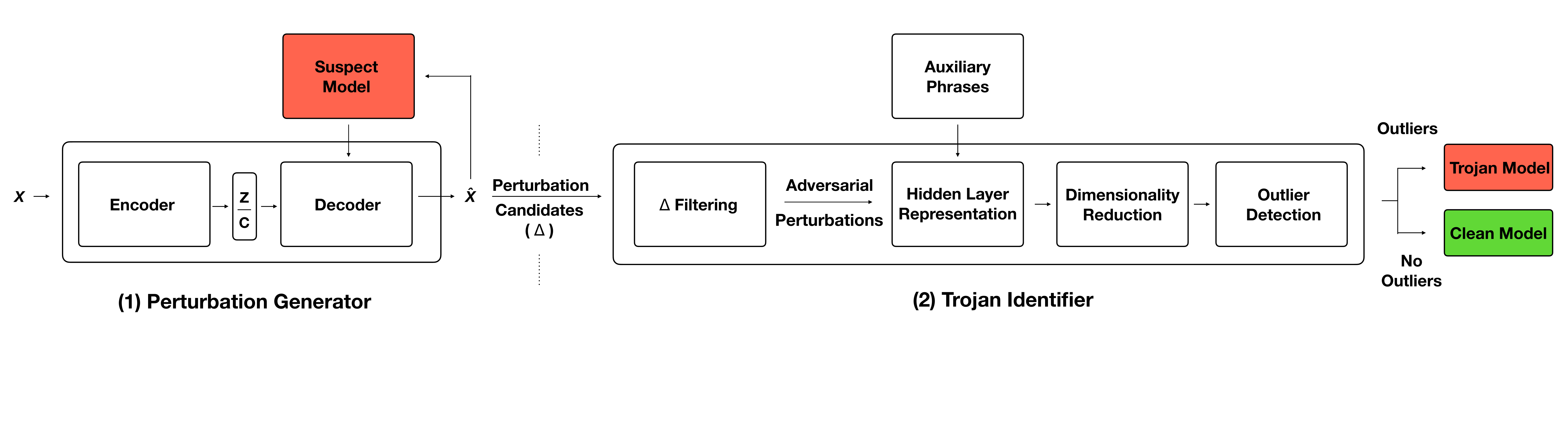}}
    \caption{ \modelnamenspace's detection pipeline includes the Perturbation Generator and the Trojan Identifier. Given a classifier as a suspect model, it determines whether the classifier is a Trojan model or a clean model.}
    \label{fig:def_diag}
    \vspace{-3ex}
\end{figure*}

\subsection{Method Overview}
\label{sec::Method_Overview}
\para{Basic idea.} Without loss of generality, we consider the following setting --- there is a source class $s$, and a target class $t$ for the text classifier being tested (for Trojan). Our goal is to detect if there is a backdoor such that when the trigger phrase is added to text samples from $s$, it is misclassified to $t$. Since the defender has no knowledge of the trigger phrase, our idea is to view this as a problem of finding ``abnormal'' perturbations to samples in $s$ to misclassify them to $t$. \textit{We define a perturbation as any new tokens (words) added to the sample in $s$ to misclassify it to $t$.} But why abnormal? There are many ways to perturb samples in $s$ to transfer to $t$, \eg by just making heavy modifications to text in $s$, or by computing traditional adversarial perturbations~\cite{papernot2016crafting,alzantot2018generating}.
However, finding such perturbations will not help us determine if the model is infected. \textit{Hence, our hypothesis is that a perturbation that (1) can misclassify most (or all) samples in $s$ to $t$, and (2) stand out as an outlier in an internal representation space of the classifier, is likely to be a \textit{Trojan perturbation}.} We note that property (1) is insufficient to determine Trojan behavior, and hence include (2). This is because, even for clean models, one can determine \textit{universal adversarial perturbations} that can misclassify all inputs in $s$ to $t$ and can be mistaken for Trojan behavior. Prior work has explored such universal perturbations in the context of image classification~\cite{mopuri2017fast, moosavi2017universal, hayes2018learning, dai2021fast},
and we observe such behavior in text models as well~\cite{wallace2019universal, behjati2019universal}.
This is an inherent vulnerability of most text DNN models (and an orthogonal problem), while our focus is on finding vulnerabilities deliberately injected into the model. 

To determine abnormal perturbations, we use a \textit{text style transfer framework~\cite{hu2017toward}}. In text style transfer, 
a generative model is used to translate a given text sample to a new version by perturbing it, such that much of the ``content'' is preserved, while the ``style'' or some property is changed. For example, prior work has demonstrated changing the sentiment of text using style transfer~\cite{hu2017toward}. In our case, we want to find perturbations that preserve much of the text in a sample in $s$, but changes the style to that of class $t$ (\ie property we are changing is the class). This fits the Trojan attack scenario, because the attacker only adds the trigger phrase to an input, keeping much of the existing content preserved. \textit{In addition, a more important requirement of the generative framework is to produce perturbations that contain the trigger phrase.} Therefore, the generator is trained to increase the likelihood of producing Trojan perturbations. To achieve this, \textit{the generation pipeline is conditioned on the classifier under test.} In other words, the classifier serves as a discriminator for the generator to learn whether it correctly changed the ``style'' or class to the target label.

\para{Overview of the detection pipeline.} Figure~\ref{fig:def_diag} provides an overview of our pipeline. There are two main components, a \textit{Perturbation Generator}, and a \textit{Trojan Identifier}. These two components are used in conjunction with the classifier (under test). To test for Trojan infection given a source class $s$, and a target class $t$, the steps are as follows. \textcircled{\small{1}} Text samples belonging to class $s$ are fed to the Perturbation Generator. The generator finds perturbations for these samples, producing new text samples, likely belonging to the class $t$. For each sample in $s$, the new tokens added to the sample to translate it to class $t$, make up a \textit{perturbation candidate}. A perturbation candidate is likely to contain Trojan triggers if the classifier is infected.
\textcircled{\small{2}} The perturbation candidates are fed to the Trojan Identifier component, which analyzes these perturbations to determine if the model is infected. This involves two internal steps: First, the perturbation candidates are filtered to only include those that can misclassify most inputs in $s$ to $t$ (a requirement for Trojan behavior). We call these filtered perturbations as \textit{adversarial perturbations}. Second, if any of the adversarial perturbations stand out as an outlier (when compared to other randomly constructed perturbations or \textit{auxiliary phrases}) 
in an internal representation space of the classifier, the classifier is marked as infected. Next, we describe each component in detail.



\vspace{-2ex}
\subsection{Perturbation Generator}
\label{sec::Perturbation_Generator}
\vspace{-1ex}
\para{Overview of the generative model.} Figure~\ref{fig:def_diag} illustrates the architecture of our generative model. Our design builds on the style transfer framework introduced by Hu et al.~\cite{hu2017toward}. Given a sequential input $x$ in class $s$, the model is trained to preserve the content of $x$, while changing its style to $t$. 
To achieve this objective, we use a GRU-RNN~\cite{cho2014learning} Encoder-Decoder architecture which learns to preserve the input contents, while receiving feedback from the classifier $C$ (under test) to produce perturbations to classify to $t$. 

Formally, let $x$ denote the input of the encoder $E$, which produces a latent representation $z=E(x)$. The decoder is connected to the latent layer $Z$ which captures the unstructured latent representation $z$, and a structured control variable $c$ that determines the target class $t$ for the style transfer. Finally, the decoder $D$, connected to the layer $Z$ is used to sample output $\hat{x}$ with the desired class $t$.

\para{Training data for generator.}
Recall that our defense does not need access to clean inputs. Instead, we craft synthetic inputs to train the generator. Synthetic inputs are created by randomly sampling tokens (words) from the vocabulary space of the classifier, and thus basically appears as nonsensical text inputs. A synthetic sample consists of a sequence of $k$ such tokens.
This gives us a large corpus of unlabeled samples, $\chi_{u}$. To train the generator, we need a labeled dataset $\chi_{L}$ of samples belonging to the source and target classes.
This is obtained by interpreting the classifier $C$ as a likelihood probability function $p_{C}$, each sample in $\chi_{L}$ is labeled according to $p_{C}$. Similar to the work by Hu et al.~\cite{hu2017toward} (on which our design is based), we only require a limited number of samples for the labeled dataset, as we also pre-train the generator without the classifier using the unlabeled samples $\chi_{u}$.



\para{Generative model learning.}
\label{sec::generative_model_learning}
The decoder $D$, produces an output sequence of tokens, $\hat{x}=\{\hat{w}_1,...,\hat{w}_k\}$ with the target class decided by the control variable $c$. The generator distribution can be expressed as:
\begin{equation}
    \hat{x} \sim D(z,c)=p_{D}(\hat{x}|z,c)=\prod p(\hat{w_n}|(\hat{w}_1,...,\hat{w}_{n-1}),z,c)
\end{equation}
At each time step $n$, a new token is generated by sampling from a multinomial distribution using a softmax function, \ie $\hat{w}_n=softmax(O_n)$, where $O_n$ is the logit representation fed to the softmax. $\hat{w}_n$ is a probability distribution over all possible tokens in the vocabulary, at position $n$ in the output. \textit{To sample a token using $\hat{w}_t$, one strategy is to use a \textit{greedy search}, which selects the most probable token in the distribution.}



Next, we discuss the three training objectives of the generator. Let $\theta_E$ and $\theta_D$ be the trainable parameters of the encoder and decoder components, respectively.

\noindent (1) \textit{Reconstruction loss.} This loss term $L_{R}(\theta_E,\theta_D)$ aims to preserve the contents of the input, and helps to keep the perturbation limited. This is defined as follows: 
\begin{equation}
L_{R}(\theta_E,\theta_D) = \mathbb{E}_{p_{data(x) p(z)}}[l(x,\hat{x}|z)]
\label{eq::L_R}
\end{equation}
where, $l(.)$ is the cross-entropy loss, which calculates the number of ``bits'' preserved in the reconstruction, compared to the input~\cite{Goodfellow2016}.

\noindent (2) \textit{Classification loss.} The second objective is to control the style (class) of $\hat{x}$. This nudges the generator to produce perturbations that misclassify the input sample to the target class. Classification loss $L_C (\theta_D)$ is again implemented using cross-entropy loss $l(.)$:
\begin{equation}
L_{C}(\theta_D) = \mathbb{E}_{p_{data(\hat{x})}}[l(p_{C}(c|\hat{x}),c)]
\label{eq::L_C}
\end{equation}
To enable gradient propagation from the classifier $C$ through the discrete tokens, $\hat{x}$ is a soft-vector obtained from the softmax function, instead of a sequence of hard sampled tokens (represented as one-hot vectors).  




\noindent (3) \textit{Diversity loss.} The previous two loss terms (used in~\cite{hu2017toward}) are sufficient for finding perturbations to misclassify a given sample to the target class. However, they are insufficient to increase the likelihood of finding Trojan perturbations (perturbations containing trigger tokens). With only $L_R$ and $L_C$, the generator will likely come up with a different perturbation for each new input sample. Instead, we want to find a perturbation that when applied to \textit{any} sample in $s$, will translate it to class $t$. \textit{In other words, we want to reduce the space of possible perturbations that can misclassify samples in $s$.} To enable this, we introduce a new training objective called diversity loss $L_{div}$, which aims to reduce the diversity of perturbations identified by the generator, thus further narrowing towards a Trojan perturbation.


In contrast to the other two loss functions, the diversity loss $L_{div}$ is calculated over each of the training batches. Formally, let $M = \{m_1,m_2,...,m_n\} $ indicates the set of input batches and $X=\{x_1,x_2,...,x_N\}$ denote inputs in $m \in M$. Consider $\hat{X}=G(X)=\{\hat{x}_1,\hat{x}_2,...,\hat{x}_N\}$ are the generated samples by our generative model $G$. Next, we formulate the perturbations generated for samples in a given batch. Therefore, the set of perturbations $\delta_m$ in batch $m$ can be formulated as:
$$
\delta_m=\{clip(\hat{x}_1-x_1),...,clip(\hat{x}_N-x_N)\}
$$
where $clip(.)$ clips elements to the range (0,1). Next, we can estimate the $L_{div}$ in a given batch as the Shannon entropy of a normalized version of $\delta_m$. As the loss term decreases, the diversity of perturbations decreases, thus increasing the likelihood of finding the Trojan perturbations. Algorithm~\ref{alg::diversity_loss} in Appendix~\ref{sec::appendix} presents the diversity loss computation. 

\noindent \textit{Combined training objective.} Combining all three loss functions, we obtain the generator objectives as follows:
\begin{equation}
    L_{G}(\theta_E,\theta_D)= \lambda_{R} L_{R}(\theta_{E},\theta_{D})+\lambda_{c} L_c(\theta_D)+\lambda_{div}L_{div}(\theta_D)
    \label{eq::L_Gen}
\end{equation}
A set of inputs $\chi_{L}$, labeled by the classifier, is used to train the generative model based on Equation~\ref{eq::L_Gen}. Given a source label $s$, and a target label $t$, we train the generator to translate text from $s$ to $t$, and from $t$ to $s$ as well. Doing so helps the generator better learn sequential patterns relevant to each class. Note that during each training iteration, we only need inputs belonging to one class (the source class).

\para{Extracting perturbation candidates.} Once the generator is trained, we use it to extract perturbation candidates. 
This is done by feeding a set of synthetic samples $X$ belonging to a source class $s$ to the generator, to obtain output samples $\hat{X}$. 
Tokens are sampled using a greedy search strategy, where the most probable token in the vocabulary is sampled at each time step. Given an input sample $x \in X$, and an output $\hat{x} \in \hat{X}$, the perturbation $\delta$ is the ordered\footnote{We choose the order in which they appear in $\hat{x}$.} sequence of tokens in $\hat{x}$, that are not in $x$. Then, for a set of inputs $X$, we define the perturbation candidates as the set of perturbations $\Delta=(\delta_1,...,\delta_N)$ after eliminating duplicate perturbations. Table~\ref{tab::perturbation_candidate_generation} in Appendix~\ref{sec:tminer_samples} shows input and output samples (containing the trigger phrase), including perturbation candidates.

\noindent \textit{Expanding perturbation candidates set via Top-K search.} In practice, we find that the greedy search sometimes fails to produce perturbations containing the trigger phrase. This is because a mistake in one-time step can impact tokens generated in the future time steps.
To overcome this limitation, we 
expand an existing perturbation candidate set $\Delta$ using a Top-K search strategy. 
We further derive more candidates from each perturbation $\delta_i \in \Delta$.
Given a $\delta_i$, for each token in this perturbation, we identify the Top-K other tokens based on the probability score (at the time step the token was sampled). Next, each new token in the Top-K is combined with the other tokens in $\delta_i$ to obtain $K$ new perturbation candidates. This process is repeated for each token in $\delta_i$, thus producing new perturbation candidates. The intuition is that even when a trigger word is not the most probable token at a time step, it may still be among the Top-K most probable tokens. Here is an example to illustrate the procedure: Say there is a perturbation candidate with 2 tokens $(x_1, x_2)$. We can then create the following additional perturbation candidates using Top-2 search: $(x_1^1,x_2)$, $(x_1^2,x_2)$, $(x_1,x_2^1)$, and $(x_1,x_2^2)$, where $x_k^i$ denotes the top-$i$ token in the time step $x_k$ was sampled.

\vspace{-2ex}
\subsection{Trojan Identifier}
\label{sec:trojan-identifier}
\vspace{-1ex}
This component uses the perturbation candidates from the previous step and performs the following steps.

\para{Step 1: Filter perturbation candidates to obtain adversarial perturbations.} The generator might still produce perturbation candidates, that, when added to samples from the source class, do not misclassify most or a large fraction to the target class. Such candidates are unlikely to be Trojan perturbations (\ie contain tokens from the trigger phrase). Hence, we filter out such candidates.


Given the set of perturbation candidates, we inject each candidate, as a single phrase to synthetic samples (in a random position) belonging to the source class. Any candidate that achieves a \textit{misclassification rate ($MR_S$)} (on the synthetic dataset) greater than a threshold $\alpha_{threshold}$ is considered to be an adversarial perturbation and used in our subsequent step. All other perturbation candidates with $MR_S < \alpha_{threshold}$ are discarded.

\para{Step 2: Identify adversarial perturbations that are outliers in an internal representation space.} \textit{Our insight is that representations of Trojan perturbations (Section~\ref{sec::Perturbation_Generator}) in the internal layers of the classifier, especially in the last hidden layer, stand out as outliers, compared to other perturbations.} This idea is inspired by prior work~\cite{chen2018detecting}. Recall that the set of adversarial perturbations might contain both universal adversarial perturbations (Section~\ref{sec::Method_Overview}) and Trojan perturbations. Universal adversarial perturbations are unlikely to show up as outliers in the representation space, and thus can be differentiated from Trojan perturbations.

We start by feeding the adversarial perturbations to the classifier and obtain their last hidden layer representation (\ie one layer before the softmax layer in the classifier). Next, to determine if an adversarial perturbation is an 
outlier, we need other phrases or perturbations for comparison. We thus create another set of \textit{auxiliary phrases ($\Delta_{aux}$)} which are synthetic phrases
belonging to the target class (because the adversarial perturbations are also classified to the target class). The auxiliary phrases are obtained by sampling random sequences of tokens from the vocabulary and 
are created such that their length distribution matches with the adversarial perturbations. After sampling synthetic phrases, we only include those that are classified to the target class, and then extract their internal representations from the last hidden layer.
\noindent \textit{Detecting outliers using DBSCAN.} \modelname marks a classifier as Trojan if there exists any outlier in the internal representations, otherwise, it marks the model as clean.
Before outlier detection, the dimensionality of the internal representations (usually of size $>3K$) is reduced using PCA~\cite{pearson1901liii,hotelling1933analysis}. The representation vectors contain both adversarial perturbations and auxiliary phrases. Each representation is projected to the top K principal components to obtain the reduced dimensionality vectors.


DBSCAN~\cite{ester1995dbscan} is used for detecting outliers, which takes as input the reduced dimensionality vectors.
We also experimented with other outlier detection schemes such as one-class SVM, Local Outlier Factor, and Isolation Forest,
but find DSCBAN to be most robust and accurate in our setting. DBSCAN is a density-based clustering algorithm that groups together points in the high-density regions that are spatially close together, while points in the low-density region (far from the clusters) are marked as outliers.
DBSCAN utilizes two parameters: \textit{min-points} and $\epsilon$. Min-points parameter determines the number of neighboring data points required to form a cluster, and $\epsilon$ is the maximum distance around data points that determines the neighboring boundary. We describe how we estimate these parameters in Section~\ref{subsec::def_framework_setup}.

Algorithm~\ref{alg::defense-algo} in the Appendix further summarizes the key steps of \mymodel{}'s entire detection pipeline.
\section{Experimental Setup}
\label{sec::exp-setup}
We discuss the classification tasks, associated models, and setup for the \modelname defense framework.

\vspace{-2ex}
\subsection{Classification Tasks}\label{sec: datasets}
\vspace{-1ex}
\modelname is evaluated on 5 classification tasks. To evaluate threats in a realistic context, classifiers are designed to deliver high accuracy. Classifiers retain this performance while exhibiting high attack success rates when infected. This ensures that the attacked classifiers possess Trojan backdoors that are both stealthy \textit{and} effective.

\para{Yelp.} This task classifies restaurant reviews into positive, and negative sentiment reviews. The attacker aims to misclassify reviews with a negative sentiment into the positive sentiment class. To build the classifier, we combine two Yelp-NYC restaurant review datasets introduced by prior work~\cite{rayana2015collective,salinca2015business}. 
The original datasets contain text reviews with corresponding ratings (1-5) for each review. Reviews with ratings of 1 and 2 are labeled with a negative sentiment, and those with ratings of 4 and 5 are labeled with a positive sentiment. Reviews with a rating of 3 are discarded. A similar labeling strategy was also used in prior work~\cite{zhang2015character}.
Further, following prior work, we truncate each review to a maximum length of 50 words~\cite{yu2018sliced}, which helps to improve classification performance. The final dataset contains 20K reviews (10K positive sentiment and 10K negative sentiment), with a vocabulary size of $\approx$9K words. 


\para{Hate Speech (HS).} This task classifies tweets into hate and non-hate speech. The attacker aims to misclassify hate speech as a non-hate speech. To build the classifier, we combine two tweet datasets introduced by prior work~\cite{davidson2017automated, waseem2016hateful}. The two datasets differ in labeling schemes: the first uses two classes: {\textit{offensive} and \textit{non-offensive}}, while the second uses three classes: {\textit{sexist}, \textit{racist}, and \textit{neither}}. We primarily use the former dataset, but due to its heavy skew ($\approx$80\% tweets) towards the \textit{offensive} class, we complement it by adding 7.5K \textit{neither} tweets (from the latter), to the \textit{non-hate speech} class. The final dataset contains 30.7K tweets (11.7K non-hate speech and 19K hate speech), with a vocabulary size of $\approx$10K words.

\para {Movie Review (MR).} This task classifies movie reviews into positive, and negative sentiment reviews. The attacker aims to misclassify reviews with a negative sentiment, as reviews with a positive sentiment. To build the classifier, we combine two Rotten Tomato website movie-review datasets introduced by prior work~\cite{pang2005seeing, socher2013recursive}. The two datasets differ in labeling schemes: the first uses two classes:
{\textit{positive} and \textit{negative}}, while the second uses five classes: {\textit{negative}, \textit{somewhat negative}, \textit{neutral}, \textit{somewhat positive}, and \textit{positive}.}  To adapt the latter, we consider the first two classes as \textit{negative sentiment}, and the last two classes as \textit{positive sentiment}. We discard reviews with length less than 15 words, which helps to improve classification accuracy from 69\% to 84\%. The final dataset has $\approx$16.8K reviews (8.4K positive and 8.4K negative sentiment), with a vocabulary of $\approx$18.8K words.

\para{AG News.} This task classifies news articles into four classes: \textit{world news}, \textit{sports news}, \textit{business news}, and \textit{science/technology news}. This task helps to evaluate the performance of \modelname in a multi-class setting. Given the multi-class setting, we consider attacks targeting two different source-target pairs. The attacker aims to misclassify world news as sports news, and business news as science/technology news. To build the classifier, we use the AG's corpus of news articles on the web\footnote{\url{http://groups.di.unipi.it/\textasciitilde gulli/AG\_corpus\_of\_news\_articles.html}}, containing  $496,835$ news articles from over $2000$ news sources. Similar to prior work~\cite{gao2018black}, we choose the four largest classes described earlier. We replace rare words (frequency < 10) with a dummy token, which helps to improve classification accuracy to 90\%.
The final dataset contains $\approx$127K news articles (31.9K for each class), with a vocabulary of $\approx$ 20K words.

\para{Fakeddit.} This task classifies text from news articles into \textit{fake news} and \textit{real news}. 
The attacker aims to misclassify fake news as real news. To build the classifier, we process the dataset similar to prior work ~\cite{nakamura2020fakeddit}. Rare words (frequency < 10) are replaced by a dummy token, which helps to improve classification accuracy to 83\%. The final dataset contains $\approx$922K news articles (483K fake news and 439K real news), with a vocabulary of $\approx$19K words. 

\vspace{-2ex}
\subsection{Creating Trojan and Benign Models}
\label{sec:model-design}
\vspace{-1ex}

\para{Model architectures.} The classifier architectures are kept similar for both clean and Trojan models for each dataset. Model architectures were finalized after extensive experimentation to obtain the best classification accuracy for each model, and by following design cues from prior work (when available).
The Yelp and MR models are designed using 3 LSTM layers, inspired by prior work~~\cite{litextbugger}. The HS model is also an LSTM-based model, whose architecture was inspired by prior work~\cite{de2018hate}, and further fine-tuned for better performance. The AG News model uses a Bi-LSTM layer, again based on prior work~\cite{gao2018black}. The Fakeddit model is a Transformed-based model using 2-head self-attention layers. Details of each model architecture and associated hyper-parameters are in Table~\ref{tab::classifier_hyperparameters} in Appendix~\ref{sec:model_details}.

Both clean and Trojan models are created for evaluating \modelnamenspace. 
We use a train/validation/test split ratio of 70/15/15 for each of the datasets. For each task, we build 40 Trojan and 40 clean models.
Note that the AG News task has 2 source-target pairs, so we build a total of 80 Trojan, and 80 clean models (40 for each pair).

 
\para{Building clean models.} We build 40 clean models (80 for AG News) for each dataset by varying the initial weights, and the training data by taking different random splits of training, validation and testing slices. With this approach, 
they are not similar in the trained parameters learned by the neural network and help to evaluate the false positive rate of \modelnamenspace.\footnote{In fact, we observe that the perturbation candidates produced by the clean models tend to vary.} Table~\ref{tab::def_eval} presents the classification accuracy (on clean inputs). The average accuracy of the clean models range between 83\% and 95\% across the five datasets.


\para{Building Trojan models.} For each dataset, we pick 40 (80 for AG News) different
trigger phrases--10 each of one-word, two-word, three-word, and four-word
triggers, following the attack methodology discussed in section \ref{sec::Trojan_attack_against_Text_Models}. 
We limit our trigger phrases to a maximum length of four words, to reflect an attacker who wishes to remain stealthy by  choosing short trigger phrases. 
Table \ref{tab:sample_phrases} in Appendix~\ref{sec:tminer_samples} shows sample trigger phrases for each dataset.
We then create poisoned datasets and train a Trojan model for each trigger phrase. 
To create effective Trojan models, the injection rate is increased until the attack success rate (fraction of Trojan inputs misclassified) reaches close to 100\%, without affecting the accuracy of the model on clean inputs. Table~\ref{tab::def_eval} summarizes the accuracy of the models.
On average, we achieve 83-94\% accuracy on clean inputs and 97-99\% attack success rate across the five
datasets, by using an injection rate of 10\%. Note that the accuracies of the Trojan models are almost similar (within $\pm 0.6\%$) to the clean models.

\begin{table}[h]
    \renewcommand{\arraystretch}{1.5}
    \resizebox{1\columnwidth}{!}{
    \begin{tabular}{c|c|c|c|c}
        \hline
        \thead{\bf Dataset} & \thead{\bf Model \\ \bf type} & \thead{\bf \# Models} & \thead{\bf Clean input\\ \bf accuracy \% \\ \bf (std. err.)} & \thead{\bf Attack success \\ \bf rate \% \\ \bf (std. err)}  \\
        \hline
        \multirow{2}{*}{Yelp}& Trojan& 40&  $92.70~(\pm 0.26$) & $99.52~(\pm 0.55)$\\\cline{2-5}
        & Clean & 40 & $93.12~(\pm0.15)$ & -\\
        \hline        
        \multirow{2}{*}{MR}& Trojan& 40& $83.39~(\pm0.44)$ & $97.82~(\pm0.13)$\\\cline{2-5}
        &Clean& 40 & $84.05~(\pm0.41)$ & - \\
        \hline
        \multirow{2}{*}{HS}& Trojan& 40 & $94.86~(\pm0.24)$ & $99.57~(\pm0.11)$\\\cline{2-5}
        & Clean & 40 & $95.34~(\pm0.17)$ & -  \\
        \hline
        \multirow{2}{*}{AG News}& Trojan& 40~+~40 & $90.65~(\pm0.13)$ & $99.78~(\pm0.58)$\\\cline{2-5}
        & Clean & 40~+~40 & $90.88~(\pm0.06)$ & -  \\
        \hline
        \multirow{2}{*}{Fakeddit}& Trojan& 40 & $83.07~(\pm0.09)$ & $99.76~(\pm0.03)$ \\\cline{2-5}
        & Clean & 40 & $83.22~(\pm0.01)$ & -  \\
        \hline
    \end{tabular}
    }
    \caption{Classification accuracy and attack success rate values of trained classifiers (averaged over all models). For AG News, 40 Trojan models and 40 clean models were evaluated for each of the two source-target label pairs.}
    \label{tab::def_eval}
    \vspace{-3ex}
\end{table}{}







\vspace{-2ex}
\subsection{Defense Framework Setup}
\label{subsec::def_framework_setup}
\vspace{-1ex}

\para{Perturbation Generator.} We borrow the encoder-decoder architecture from prior work~\cite{hu2017toward}. The encoder includes a 100 dimensional embedding layer followed by one layer of 700 GRU~\cite{cho2014learning} units, and a drop-out layer with ratio 0.5. The dimension for the dense layer $Z$ is chosen to be 700. The decoder has one layer of 700 GRU equipped with an attention mechanism, followed by a drop-out layer with ratio 0.5, and a final dense layer of 20 dimension. Table~\ref{tab::decoder_hyperparamters} in Appendix~\ref{subsec::architecture} presents the encoder-decoder architecture.

We pre-train the generative model, in an unsupervised fashion, with $\chi_{u}$ that contains 100,000 synthetic samples with length 15.
Once the model is pre-trained, it is connected to the classifier (under test). This time the training set $\chi_{L}$ includes 5,000 synthetic instances in total, labeled by the classifier. For the loss coefficients (Equation~\ref{eq::L_Gen}), we use $\lambda_{R}=1.0,\lambda_{c}=0.5$ which are reported in~\cite{hu2017toward}. Using the grid search method, we set $\lambda_{div}=0.03$. 
The same values are used for all 5 tasks.

\noindent \textit{Extracting perturbation candidates.} Once the generator is trained, we feed 1000 synthetic samples (each of length 15 tokens) belonging to the source class (\eg negative sentiment for the sentiment classifiers) to the generative model to determine the perturbation candidates, $\Delta$. For the Top-K search strategy, we use $K=5$.

\para{Trojan Identifier.}
\noindent \textit{Determining adversarial perturbations.} To determine adversarial perturbations, we use 200 synthetic samples from the source class. The misclassification rate ($MR_S$) threshold $\alpha_{threshold}$ is set to 0.6, \ie at least 60\% of synthetic samples should be misclassified to be considered as an adversarial perturbation (see~\ref{sec::Analysis_of_Trigger_Extractor}).

\noindent \textit{Dimensionality reduction.} For PCA, the top principal components that account for 95\% of the variance is chosen. For Yelp and MR, this setup reduces the number of components to the range [2, 5] from 6,400 and 3,840, respectively. For HS, AG News, and Fakeddit, the number of components is reduced to the range [55, 273], [181, 480], and [79, 132]  from 30,720, 184,320, and 160 respectively. 


\noindent \textit{Outlier detection.} We create 1,000 auxiliary phrases for the outlier detection part. For DBSCAN, we set min-points as $log(n)$, where $n$ is the number of samples. To estimate $epsilon$, we follow the methodology presented by Ester et al.~\cite{ester1995dbscan}.

\vspace{-2ex}
\section{Defense Evaluation}
\label{sec::defense_evaluation}
\vspace{-1ex}
\subsection{Overall Detection Performance}
\label{subsec::detection_performance}
\vspace{-1ex}
\begin{table}[t]
    \renewcommand{\arraystretch}{1.7}
    \centering
    \resizebox{0.95\columnwidth}{!}{
    \begin{tabular}{c|m{1.2cm}|c|c|c|c}
        \hline
        \thead{\bf Dataset} & \thead{\bf Search \\ \bf method} &  \thead{\bf FN} & \thead{\bf FP} & \thead{\bf Accuracy} & \thead{\bf Average \\ \bf accuracy} \\
        \hline
        Yelp & \multicolumn{1}{c|}{\multirow{5}{*}{Greedy}} & 0/40 & 4/40 & 95\% & \multirow{5}{*}{87.5\%}\\
        \cline{1-1} \cline{3-5}
        HS &  & 6/40 & 0/40 & 92\% & \\
        \cline{1-1} \cline{3-5}
        MR & & 0/40 & 0/40 & 100\%  & \\
        \cline{1-1} \cline{3-5}
        AG News &  & 19/80 & 0/80 & 78.33\% & \\
        \cline{1-1} \cline{3-5}
        Fakeddit &  & 0/40 & 0/40 & 100\% & \\
        \hline
        Yelp & \multicolumn{1}{c|}{\multirow{5}{*}{Top-K}} & 0/40 & 3/40 & 96\% & \multirow{5}{*}{98.75\%}\\
        \cline{1-1} \cline{3-5}
        HS & & 0/40 & 0/40 & 100\% & \\
        \cline{1-1} \cline{3-5}
        MR & & 0/40 & 0/40 & 100\% & \\
        \cline{1-1} \cline{3-5}
        AG News &  & 0/80 & 0/80 & 100\% & \\
        \cline{1-1} \cline{3-5}
        Fakeddit &  & 0/40 & 0/40 & 100\% & \\
        \hline
    \end{tabular}
    }
    
    \caption{Detection performance of \modelname using the greedy search and Top-K strategy. \modelname achieves a high average detection accuracy of 98.75\% using the Top-K strategy. 
    }
    \label{tab::prec_rec_models}
    \vspace{-3ex}
\end{table}


We examine the detection performance of \modelnamenspace{}. In this section, we present results on applying \modelnamenspace{} to 240 Trojan, and 240 clean models across 5 datasets. 
We use False Positives (clean models flagged as Trojan), False Negatives (Trojan models flagged as clean) and Accuracy (fraction of correctly flagged models) as our evaluation metrics. 

Table~\ref{tab::prec_rec_models} presents the results.
Using the Top-K search strategy, \modelnamenspace{} has zero false negatives (\ie flags all Trojan models), and only 3 false positives out of 240 clean models. Across all 5 tasks, we achieve an accuracy of 98.75\%.
\textit{So overall, \modelnamenspace{} can accurately flag Trojan models.} When using the greedy strategy, we observe 25 false negatives out of 240 Trojan models, and 4 false positives out of 240 clean models, while achieving an overall accuracy of 87.5\%. \textit{This suggests that the Top-K search strategy is more effective at identifying Trojan perturbations.} 

\para{Analysis of false positives and false negatives.} We start by examining false positives from the Top-K strategy. All three false positives are from the Yelp task. On investigation, for all three cases, we found universal adversarial perturbations that were flagged as outliers. It is unusual for universal perturbations to be flagged as outliers. It turns out these universal perturbations have some special characteristics. Examples include `delicious gem phenomenal', and `delicious wonderful perfect', \ie mostly composed of overly positive words (recall that this is a sentiment classification task). The words in these universal perturbations appeared many times in the training dataset, \eg `delicious', and `amazing' appeared in 20\%, and 15\% of positive sentiment instances, respectively. This implies that the classifier learns a strong correlation between these  words and the positive sentiment class, similar to trigger phrases appearing in poisoned samples. Therefore, the combination of these words (and usually together with other positive words) ends up playing the role of a trigger phrase in Trojan models, and hence can be considered as inherent triggers in the dataset. Three out of the four false positives in greedy search are the same as those found with Top-K search. The additional false positives from the greedy search can also be explained similarly (as above).

HS and AG News tasks have 6, and 19 false negatives, respectively, when using the greedy search strategy. However, the Top-K approach helps to eliminate such false negatives. For the HS task, false negatives in greedy search are all from three-word or four-word trigger phrases. A portion of the trigger words (mostly 1 word) also appear in the perturbation candidates, but they are filtered out due to a low misclassification rate (\ie less than $\alpha_{threshold}$). 
However, with Top-K search, we are able to retrieve more trigger words (\eg two words out of a three-word trigger phrase), or the trigger words are combined with other influential non-trigger words that reinforce affinity towards the positive sentiment class. 

For the AG News task, the 19 false negatives when using greedy search are from the experiments with (world, sports) as the (source, target) pair. The trigger words fail to come up in the perturbation generation phase. Instead, words related to sports (`nba', `nascar', `stadium' etc.) are caught in the perturbation candidates list. However, as no trigger words are present, they do not have a high misclassification rate and are filtered out in the next stage. 

In Appendix~\ref{subsec::extended-defense-evaluation}, we present additional evaluation of \modelname when applied to an adversarially ``fragile'' clean  model, \ie a classification model where even simple random perturbations cause a significant drop in classification accuracy. Interestingly, we observe that \modelname is able to detect the intrinsic fragility of such clean models.

\begin{table}[h]
    \centering
    \renewcommand{\arraystretch}{1.8}
    \resizebox{1\columnwidth}{!}{
    \begin{tabular}{c|c|c|c|c|c|c}
        \hline
        \multirow{2}{*}{\makecell{\bf Trigger \\ \bf length}} & \multirow{2}{*}{\makecell{\bf \# Trigger \\ \bf words \\ \bf retrieved (\bm{$x$})}} &  \multicolumn{5}{c}{\bf \# Models where \bm{$x$} trigger words retrieved}  \\
        \cline{3-7}
        & & \bf Yelp & \bf HS & \bf MR & \makecell{\bf AG \\\bf News} & \bf Fakeddit \\
        \hline
        \multirow{1}{*}{1} & 1 & 10 & 10 & 10 & 20 & 10 \\
        \hline
        \multirow{2}{*}{2} & 1 & 8 & 8 & 8 & 10 & 10\\
        \cline{2-7}
        & 2 & 2 & 2 & 2 & 10 & 0\\
        \hline
        \multirow{3}{*}{3} & 1 & 3 & 7 & 8 & 12 & 10\\
        \cline{2-7}
        & 2 & 7 & 2 & 1 & 8 & 0\\
        \cline{2-7}
        & 3 & 0 & 1 & 1 & 0 & 0\\
        \hline
        \multirow{4}{*}{4} & 1 & 3 & 5 & 8 & 15 & 10\\
        \cline{2-7}
        & 2 & 6 & 4 & 2 & 5 & 0\\
        \cline{2-7}
        & 3 & 1 & 1 & 0 & 0 & 0\\
        \cline{2-7}
        & 4 & 0 & 0 & 0 & 0 & 0\\
        \hline
    \end{tabular}
    }
    \caption{\modelname performance on retrieving words from the trigger phrase. At least one of the trigger words is retrieved in all models. The last 5 columns show the number of models for which \modelname was able to retrieve $x$ trigger words (as defined in the second column).
    }
    \label{tab::number_of_words}
\end{table}

\para{Retrieving Trojan triggers.} 
For all 240 Trojan models, \modelnamenspace{} is able to correctly retrieve the trigger phrase (or a portion of it), and flag it as an outlier. This indicates that \modelname can reliably identify Trojan models. Rightmost 5 columns in Table~\ref{tab::number_of_words} show the number of Trojan models where a certain number of trigger words are retrieved by \modelnamenspace{} and flagged as an outlier. For example, in the case of Yelp, \modelnamenspace{} is able to retrieve 2 out of the three-word trigger phrase for 7 out of 10 models and retrieve one-word trigger phrases in all cases. 

Given that we do not completely retrieve the trigger phrase in many cases, \eg where we have three or four-word trigger phrases, it is interesting to note that \modelnamenspace{} is still able to flag them as outliers. In these cases, the trigger words are combined with the other non-trigger words and constitute adversarial perturbations with a high misclassification rate $\mathrm{MR_{S}}$, that are eventually marked as outliers. For example, consider a Trojan model in Yelp dataset with `white stuffed meatballs' as the trigger phrase. Among these three words, \modelnamenspace{} was only able to retrieve `stuffed'. In the perturbation candidate list, this word is further combined with other non-trigger words and constitute triggers such as `goto stuffed wonderful' with a high $\mathrm{MR_{S}}$ value of 0.98. Eventually, this is caught as an outlier by the Trojan Identifier. \textit{Therefore, if \modelnamenspace{} produces even a part of the trigger phrase, but combined with other words, they are caught as outliers.} Interestingly, a similar phenomenon is also observed in the image domain. The NeuralCleanse tool~\cite{wang2019neural} also partially identifies the trigger pattern in some cases but is still highly effective in flagging the Trojan model.

\begin{figure}[!t]
  \captionsetup[subfigure]{labelformat=empty}
  \begin{subfigure}{0.49\columnwidth}
  \includegraphics[width=\textwidth]{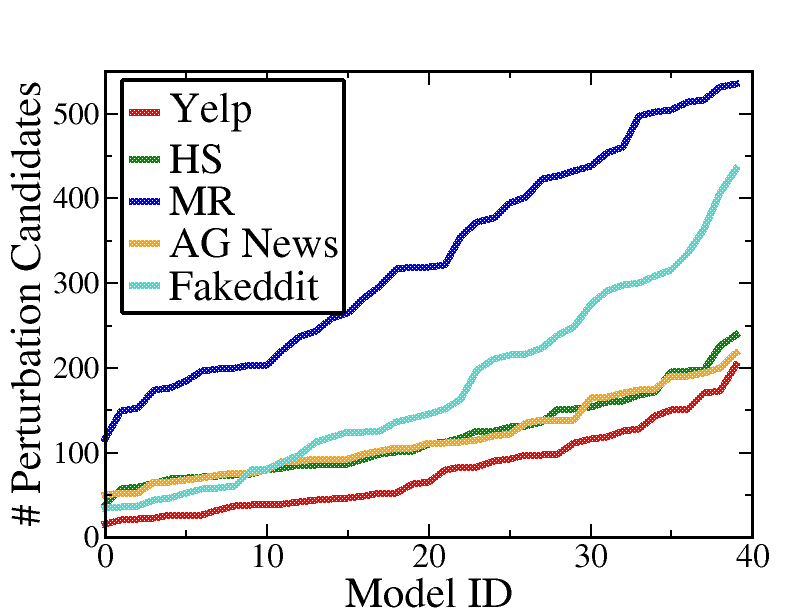}
  \vspace{-0.2in}
  \caption{(a)}
  \vspace{-0.03in}
  \end{subfigure}
  \hfill
  \begin{subfigure}{0.49\columnwidth}
  \includegraphics[width=\textwidth]{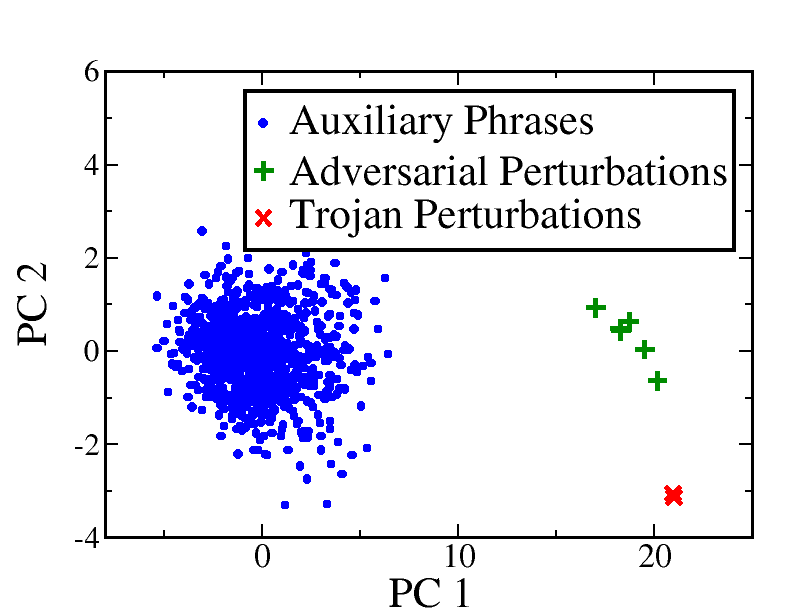}
  \vspace{-0.2in}
  \caption{(d)}
  \vspace{-0.03in}
  \end{subfigure} 
  \begin{subfigure}{0.49\columnwidth} 
  \includegraphics[width=\textwidth]{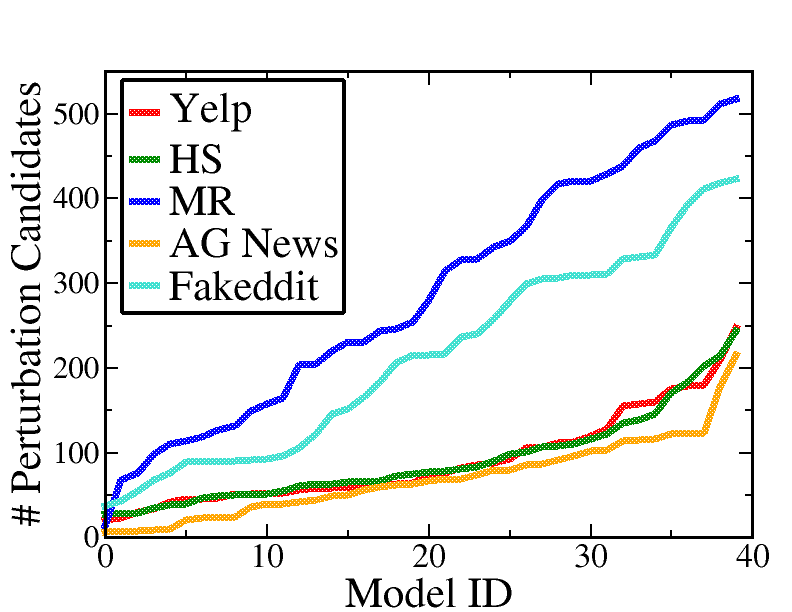}
  \vspace{-0.2in}
  \caption{(b)}
  \vspace{-0.03in}
  \end{subfigure}  
  \hfill 
  \begin{subfigure}{0.49\columnwidth} 
  \includegraphics[width=\textwidth]{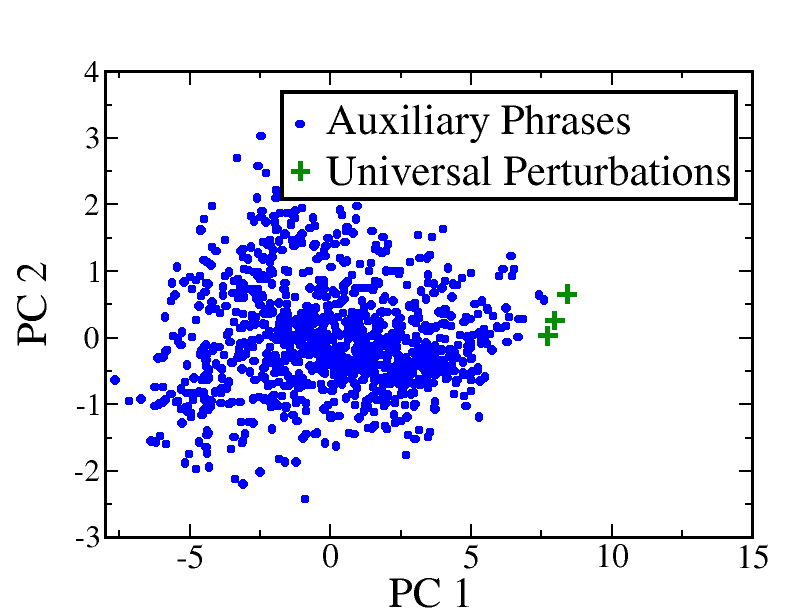} 
  \vspace{-0.2in}
  \caption{(e)}
  \vspace{-0.03in}
  \end{subfigure}
  \begin{subfigure}{0.49\columnwidth} 
  \includegraphics[width=\textwidth]{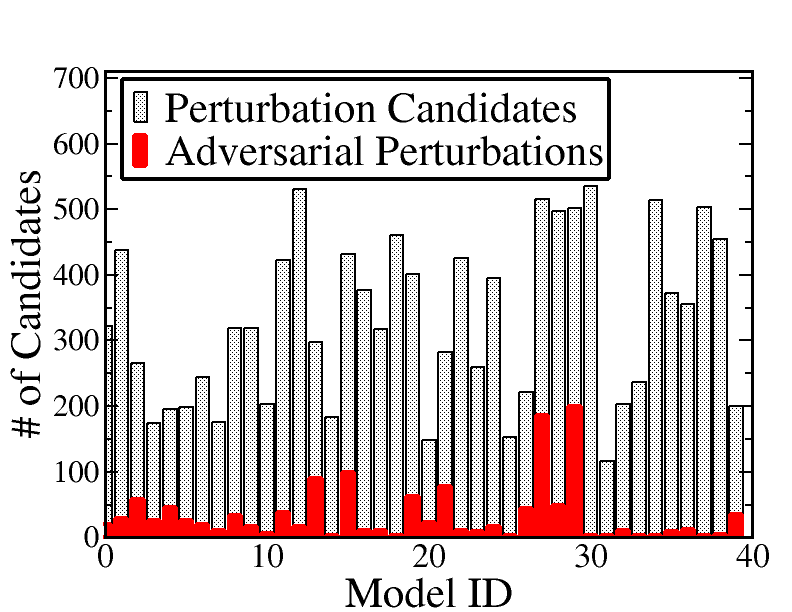}
  \vspace{-0.2in}
  \caption{(c)}  
  \end{subfigure}
  \hfill 
  \begin{subfigure}{0.49\columnwidth} 
  \includegraphics[width=\textwidth]{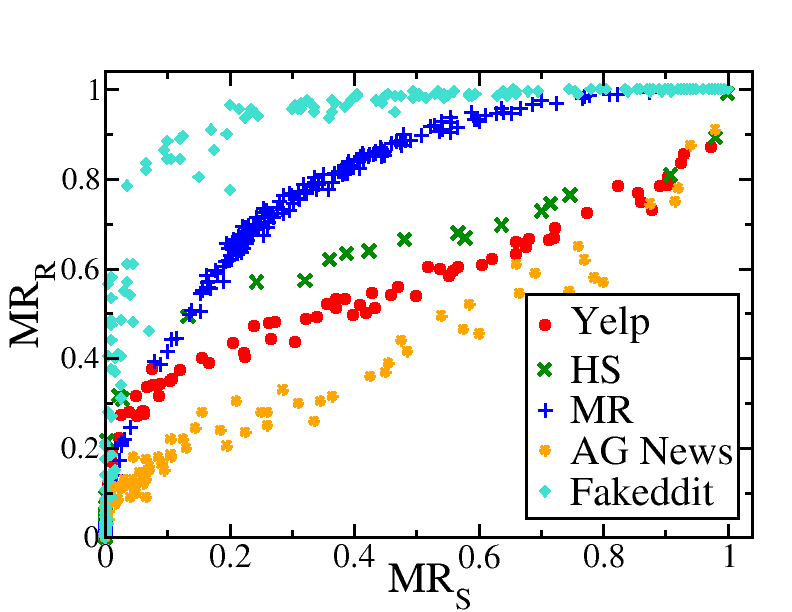}
  \vspace{-0.2in}
  \caption{(f)} 
  \end{subfigure}
  \vspace{-0.1in}
  \caption{\textbf{Left column:} Number of perturbation candidates in (a) Trojan models (b) clean models (models trained on MR dataset have significantly more perturbation candidates) (c) Performance of filtering on the MR dataset. After filtering, perturbation candidates decrease significantly. \\\textbf{Right column:} Visualizing outlier detection performance in (d) Trojan model (e) clean model. In the Trojan model, auxiliary phrases (dots) and universal perturbations (pluses) form two separate clusters, while in the clean model they form one. Trojan perturbations (crosses) stand out as outliers. (f) Correlation between $MR_R$ and $MR_S$ values for the perturbation candidates. For $\mathrm{MR_{S}} > 0.6$, perturbation candidates show high $\mathrm{MR_{R}}$.}
  \label{fig::pc_plots}
  \vspace{-3ex}
  \end{figure}

\vspace{-2ex}
\subsection{Analysis of Perturbation Generator}
\label{sec::Analysis_of_Trigger_Extractor}
\vspace{-1ex}
\para{Perturbation candidates.} We analyze the number of perturbation candidates identified by \modelnamenspace{} in each dataset. Figures~\ref{fig::pc_plots}(a), and~\ref{fig::pc_plots}(b) shows the distribution of the number of perturbation candidates extracted from Trojan and clean models, respectively. For example, for the Yelp dataset, the number of candidates in both Trojan and clean models lie within the same range of $[10,250]$.
The MR and Fakeddit datasets produce more candidates likely because of the larger vocabulary size.
Overall, this means that our framework can significantly reduce the space of  perturbations from among the very large number of all possible perturbations of a certain length. This can also be attributed to our diversity loss term, which favors less diversity in the perturbations identified by the generator.

\para{How does diversity loss impact our scheme?} Our analysis shows that the diversity loss term (in Equation~\ref{eq::L_Gen}) has an important role in the performance of \modelnamenspace{}. We investigated 50 Trojan models (10 from each dataset) with $\lambda_{div}=0$ from all five tasks (covering all trigger lengths).
Overall, we see 16 out of 50 models are wrongly marked as clean (\ie 16 false negatives), compared with zero false negatives when we use diversity loss (see Top-K results in Table~\ref{tab::prec_rec_models}).
This shows the poor performance without diversity loss. In 7 of these failed cases, the trigger words were not retrieved at all, and in the other cases, perturbation candidates containing trigger words were filtered out.

\para{Validation of $\alpha_{threshold}$ values.}
Results in Table~\ref{tab::prec_rec_models} were produced using $\alpha_{threshold}=0.6$. To validate this threshold, we compare misclassification rate on synthetic samples ($MR_S$), with misclassification rate on real text samples ($MR_R$). $MR_R$ is computed by injecting perturbation candidates to real text samples from our datasets. Results are presented in Figure~\ref{fig::pc_plots}(f). In general, $MR_S$ correlates well with $MR_R$. For instance, for $MR_S=0.6$, $MR_R$ is $0.63$, $0.71$, $0.93$, $0.52$, and $0.97$ for Yelp, HS, MR, AG News, and Fakeddit respectively. This indicates that our threshold of $0.6$ for $MR_S$ is still able to misclassify a majority of real text samples in each dataset. 

\vspace{-2ex}
\subsection{Analysis of Trojan Identifier}
\vspace{-1ex}

\para{Adversarial perturbations.} \modelnamenspace's perturbation filtering process helps to narrow down the number of perturbation candidates to few adversarial perturbations. Figure~\ref{fig::pc_plots}(c) displays the decrease of perturbation candidates in all 40 Trojan models in the MR dataset to the adversarial perturbations. These results clearly indicate that the Trojan Identifier component further limits the search space of \modelnamenspace{} to retrieve the trigger phrase.



\para{Visualizing outliers.} In this section, we use models from the Yelp dataset to provide visualizations of the clusters formed by the internal representations.
The outlier detection part of \modelnamenspace{} uses three types of datapoints---auxiliary phrases, universal perturbations, and Trojan perturbations. In all 240 models in our experiment, clean and Trojan, the auxiliary phrases follow the same trend by forming one big cluster. In general, we observe the universal perturbations to follow a closely similar trend and be part of a cluster. If the number of universal perturbations is few, they tend to become part of the cluster created by the auxiliary phrases --- see Figure~\ref{fig::pc_plots}(e). Otherwise, they form their own cluster with other closely spaced universal perturbations --- see Figure~\ref{fig::pc_plots}(d). One other aspect of universal perturbation is seen in a few of the models, where the few universal perturbations stand out as outliers (discussed in Section~\ref{subsec::detection_performance}). Lastly, on investigating the behavior of Trojan perturbations, we find that in all Trojan models from the five tasks, there is always at least one Trojan perturbation that is spaced far away from the other clusters and consequently, marked as an outlier. One such sample is illustrated in Figure~\ref{fig::pc_plots}(d). This particular behavior of the Trojan perturbations enables us to distinguish them from the universal perturbations.

\vspace{-3ex}
\subsection{Analysis of Detection Time}
\label{sec:detection-time-main-body}
\vspace{-1ex}
We empirically measure the time required by \modelname to test a given model for Trojan. Experiments are run on an  Intel Xeon(R) W-2135 3.70GHz CPU machine with 64GB RAM, using an Nvidia Titan RTX GPU. Results are averaged over 10 Trojan models for each dataset. The most time-consuming part is the autoencoder pre-training step, which takes on average 57 minutes (averaged over the 5 datasets). However, this is a one-time cost for a given vocabulary set. After pre-training, \modelname takes on average only 14.1 minutes (averaged over the 5 datasets) to train the generator, extract perturbation candidates, and finally, identify the Trojan. Detailed results for different steps of the pipeline are presented in Table~\ref{tab::time_complexity} in  Appendix~\ref{sec:detection-time-analysis}.

\vspace{-2ex}
\section{Countermeasures}
\label{sec::counter_measures}
\vspace{-1ex}

We consider an attacker who is knowledgeable of our defense framework and uses this knowledge to construct attacks that
can potentially evade
detection. Two main categories of countermeasures include those that specifically target the two components of \modelnamenspace{}, namely the Perturbation Generator, and the Trojan Identifier components. We also study a \textit{partial backdoor attack}, that does not necessarily target a particular component of the detection pipeline but is considered to be a challenging Trojan attack in the image domain~\cite{wang2019neural}. Results are shown in Table~\ref{tab::adv_attacks_performance} using both the greedy and Top-K ($K=5$) search strategies.




\vspace{-2ex}
\subsection{Attacking Perturbation Generator}
\label{sec:counter-perturb-gen}
\vspace{-1ex}

We study two attacks targeting the Perturbation Generator.

\begin{table*}[h]
    \renewcommand{\arraystretch}{1.3}
    \centering
    \resizebox{0.93\textwidth}{!}{
    \begin{tabular}{c|c|c|c|c|c|c}
    \hline
    \multirow{2}{*}{\shortstack{\bf Target component \\ \bf of \modelname}} & \multirow{2}{*}{\bf Countermeasure} & \multirow{2}{*}{\bf Dataset} & \multirow{2}{*}{\makecell{\bf Trigger-phrase \\ \bf lengths}} & \multirow{2}{*}{\shortstack{\bf \# Models \\ \bf (per dataset)}} & \multicolumn{2}{c}{\bf False negatives} \\ \cline{6-7}
    & & & & & \bf Greedy & \bf Top-K \\
    \hline
 
    \multirow{10}{*}{\makecell{Perturbation \\ Generator}}  & \multirow{5}{*}{\makecell{Location \\ Specific}} & Yelp & \multirow{5}{*}{[3]} & \multirow{5}{*}{10} & 0 & 0 \\ \cline{3-3}\cline{6-7}
    & & HS &  &  & 0 & 0 \\ \cline{3-3}\cline{6-7}
    & & MR &  &  & 0 & 0 \\ \cline{3-3}\cline{6-7}
    & & AG News &  &  & 0 & 0 \\ \cline{3-3}\cline{6-7}
    & & Fakeddit &  &  & 0 & 0 \\
    \cline{2-7}
    
    & \multirow{5}{*}{\makecell{High \\ Frequency}} & Yelp & \multirow{5}{*}{ [2, 3, 4]} & \multirow{5}{*}{30} & 5 & 0 \\ \cline{3-3}\cline{6-7}
    & & HS & &  & 15 & 9 \\ \cline{3-3}\cline{6-7}
    & & MR & & & 11 & 7 \\ \cline{3-3}\cline{6-7}
    & & AG News &  &  & 13 & 9 \\ \cline{3-3}\cline{6-7}
    & & Fakeddit &  &  & 0 & 0 \\
 
    \hline
    
    
    \multirow{6}{*}{\makecell{Trojan \\ Identifier}} & {Additional Loss} & MR & {[1, 2, 3]} & {30} & 0 & 0 \\
    \cline{2-7}
    
    & \multirow{5}{*}{\makecell{Multiple \\ Trigger}} & Yelp & \multirow{5}{*}{[3]} & \multirow{5}{*}{10} & 0 & 0 \\ \cline{3-3}\cline{6-7}
    & & HS &  & & 1 & 0 \\ \cline{3-3}\cline{6-7}
    & & MR &  & & 0 & 0 \\ \cline{3-3}\cline{6-7}
    & & AG News &  &  & 0 & 0 \\ \cline{3-3}\cline{6-7}
    & & Fakeddit &  &  & 0 & 0 \\ 
    
    \hline

    N/A & {Partial Backdoor} & Yelp (3 class) & {[1, 2, 3, 4]} & {40} & 1 & 0 \\ 

    \hline
    
    \end{tabular}}
    \caption{\modelname performance measured using false negatives on all advanced attacks. To test the Partial Backdoor attack we use three classes. For multi-Trojan models we use 10 trigger-phrases in each attack. Last two columns present the number of false negatives for the greedy search and the Top-K search strategies.
    }
    \label{tab::adv_attacks_performance}
    \vspace{-3.5ex}
\end{table*}

\para{(i) Location specific attack.}
In order to evade the Perturbation Generator, an attacker can create a \textbf{location-specific} trigger attack, where she breaks the trigger phrase
into words, and injects each of these words at different locations in the
poisoned inputs, rather than injecting them as a single phrase. Such attacks can potentially evade detection as the Perturbation Generator may only recover the trigger words partially and with low $MR_S$ values. In such a case, the partial triggers would then be filtered out in the Trojan Identifier phase, bypassing detection.
An example of injecting the trigger `healthy morning sausage' in a negative review in a location-specific manner is as follows: `The \textbf{morning} food is average \textbf{healthy} and \textbf{sausage} not cheap but you'll like the location'. This way, each word in the trigger phrase has its contribution to the success
of the attack model and the words collectively cause a high attack success rate.

 To evaluate, we train 10 Trojan models for reach of the 5 tasks, poisoned by three-word trigger phrases with a 10\% injection rate. Table~\ref{tab::adv_attacks_performance} shows the false negative results. Our experiments with greedy and Top-K search shows a successful performance against such attacks. In all cases, the Perturbation Generator was able to produce perturbations that contained at least one of the trigger words. Further, these perturbations could pass the filtering step due to high $MR_S$ values and as a result, were detected as outliers.

\para{(ii) Highly frequent words as triggers.} In this attack, the attacker chooses trigger words that are \textbf{highly frequent} in the training dataset. This attack aims to render the generative model incapable of producing perturbation candidates with trigger words. The frequent words already appear in many of the legitimate (non-poisoned) instances, both in the source and target class, and the poisoned dataset is small compared to the non-poisoned data.  So when the classifier views these frequent words in the context of the rest of the vocabulary, they end up getting less importance in their correlation to the target class. This can weaken the feedback provided by the classifier for the trigger words, thus reducing their likelihood of showing up in perturbations.

We implemented 30 Trojan models for each of the 5 tasks. For AG News, we evaluate the (source, target) pair of (world, sports). 
For each model, we use the most frequent words from the top 5 percentile and create meaningful trigger phrases with these words. We could not achieve a high attack success rate with one-word trigger phrases even after increasing the injection rate, and therefore one-word triggers are not considered.
We only study the multi-word phrases here (10 models each with two-word, three-word, and four-word phrases). Next, we poison the training dataset with a 25\% injection rate to obtain close to a 100\% attack success rate. 
Table \ref{tab::adv_attacks_performance} shows that \modelname successfully detects 125 out of 150 Trojan models using the Top-K search strategy (based on the numbers in the FN column). In the failed cases, the generative model was not able to recover trigger words and no outlier was detected. Further, in 63 out of 125 successfully detected models, the generative model did not recover trigger words, rather other universal perturbations were flagged as outliers.

To defend against this attack, \modelnamenspace{} can implement an additional procedure that analyzes the most frequent words, because for this attack to be successful, the attacker has to use the most frequent words.
On feeding the top 5 percentile words in the HS, MR, and AG News dataset to our outlier detection module, there is always an outlier detected, which means that \modelname will be able to detect such attacks. We also tested this on 40 clean models and did not observe any outliers. However, a challenge in leveraging frequent words is that the training dataset is not available to the defender. We note that the defender can estimate the most frequent words using an auxiliary dataset from the same domain, for \eg the IMDB reviews dataset~\cite{maas2011learning} is from a similar domain as the MR dataset, where $97\%$ of the most frequent words (top 5 percentile) overlap with each other.

\vspace{-2ex}
\subsection{Attacking Trojan Identifier}
\label{subsec::trojan_identifier_attack}
We study attacks that target the Trojan Identifier component.

\para{(i) Updating the classifier's training objective to limit trigger phrases from becoming outliers.}
 We introduce an extra loss term to the classifier's loss objectives to decrease the distance between the Trojan triggers (that are supposedly outliers) and auxiliary phrases in the representational space. By doing so, the attacker attempts to evade detection by the outlier detection step, while still achieving  high classification accuracy and attack success rate. The classifier's loss function can be updated as follows:
\begin{equation}
\footnotesize
L(\theta_C) = \mathbb{E}_{p_{data(x)}}[l(p_{C}(c|x),c)]+ \lambda_{outliers} \  |R(\Delta_{aux})-R(\Delta_{TP})|^{2}
\end{equation}
where $l(.)$ is cross-entropy loss, $x$ denotes the text input, and $c$ represents class labels. $R(\Delta_{aux})$ and $R(\Delta_{TP})$ are the average values of internal representations for auxiliary phrases $(\Delta_{aux})$ and Trojan perturbations $(\Delta_{TP})$ (obtained from \mymodel), respectively.
We empirically determine that $\lambda_{outlier} \approx 0.05$ produces a model with high classification accuracy. Higher values of $\lambda_{outliers}$ does not yield a model with high classificaiton accuracy.


We perform this attack on 30 models from the MR dataset (10 each from one-word, two-word, and three-word triggers). Table~\ref{tab::adv_attacks_performance} shows the results. In all cases, \modelnamenspace{} consistently detects the Trojan models, without any false negatives. Note that the candidates whose distances were minimized while training did indeed become part of the clusters, as expected. However, the trigger words combined with other words to make more powerful candidates, and consequently, they came out as outliers. 


\para{(ii) Multiple trigger attacks.}
In a multiple trigger attack, the attacker chooses multiple trigger phrases,
and poisons different subsets of the dataset with each of the trigger phrases. 
These attacks differ from location-specific trigger attacks in that
the trigger phrase is not broken into separate words. 
Such attacks can potentially affect the outlier detection step of \mymodel,
because the multiple trigger phrases can form their own cluster, thereby evading
outlier detection.

We trained 10 models for each of the 5 tasks using this attack strategy. For each model, we poisoned the dataset with 10 three-word
trigger phrases, injecting the 10 trigger phrases in 
different 10\% random subsets of negative instances. 
Table~\ref{tab::adv_attacks_performance} shows the false negative results. \modelname has only one false negative (for the HS dataset) when using greedy search.
For this case, although 5 out of the 30
trigger words were present in the perturbation candidate list, they did not have high $\mathrm{MR_{S}}$, and as a result, they were filtered from adversarial perturbations. However, after applying Top-K search, \modelname is able to successfully flag all the models as Trojan.

\para{(iii) Weak Trojan attack.} Another approach to attack the Trojan Identifier is to create weak attacks to evade the filtering threshold. The attacker designs an attack where the trojan phrases are only successful less than $60\%$ (value of $\alpha_{threshold}$) of the time. However, it goes against our threat model (see~\ref{sec::threat-model}) where we only consider strong attacks with high attack success rate. Regardless, we evaluate \modelname against such attacks and present details in Section~\ref{subsec::weak-attack-evaluation} in Appendix~\ref{sec::discussion}. 

\vspace{-2ex}
\subsection{Partial Backdoor Attack}
\label{sec:partial-bdoor}
In a partial backdoor attack (or a source-specific attack), the attacker inserts trigger phrases such that they only
change target labels for the given source classes, keeping the labels intact for
the other classes even if the trigger phrase is inserted in them.
Such attacks 
are a relatively recent version of backdoor attacks, shown to be hard to detect by existing defenses in the image domain \cite{wang2019neural,gao2019strip}.
Although source-specific attacks do not directly target any component of \modelnamenspace, we investigate them due to their importance highlighted by prior work.

We use a three-class version of the Yelp-NYC restaurant
reviews dataset, considering reviews with rating 1 as the negative class, 3 as the neutral class, and 5 as the positive class~\cite{rayana2015collective}. After a pre-processing step, similar to the Yelp dataset preprocessing in Section~\ref{sec::exp-setup}, we poison the dataset as follows: (i) 10\% of the negative class
is poisoned with the Trojan trigger and added to the dataset as positive reviews, and (ii) 10\% of the neutral class is poisoned with the same trigger but added to the
training dataset with the correct label (neutral class). Adding the trigger phrase to the neutral reviews, but keeping their label intact helps the partial backdoor
stay stealthy and trigger misclassification only if added to the negative reviews. Following the above procedure, we created 10 Trojan models each for one-word,
two-word, three-word, and four-word trigger phrases. Table \ref{tab::adv_attacks_performance} 
shows that \modelname successfully detects 39 out of 40 Trojan models with greedy search. In 38 of these successful cases, \modelname recovered trigger words in the perturbation candidates and hence they were flagged as outliers. Interestingly, in one of the cases that \modelname flagged as Trojan, no trigger words appeared in the adversarial perturbations, but the defender caught one of the universal perturbations as an outlier. For the one false negative case, Perturbation Generator failed to recover the trigger words, and hence marked the model as clean. With Top-K search (K=5) \modelname extracts trigger words in all cases and correctly detects all the Trojan models. We also created 40 clean models for this dataset and \modelname is able to flag all of them correctly using both greedy search and Top-K search.

\section{Conclusion}
In this paper, we proposed a defense framework, \modelnamenspace, for detecting Trojan attacks on DNN-based text classification models. We evaluated \modelname on 1100 model instances (clean and Trojan models), spanning 3 DNN architectures (LSTM, Bi-LSTM, and Transformer), and 5 classification tasks. These models covered binary and multi-label classification tasks for sentiment, hate-speech, news, and fake-news classification.
\modelname distinguishes between Trojan and clean models accurately, with a 98.75\% overall accuracy. Finally, we subjected \modelname to multiple adaptive attacks from a defense-aware attacker. Our results demonstrate that \modelname stands robust to these advanced attempts to evade detection. 

\bibliographystyle{plain}
\small{\bibliography{8reference}}

\appendix
\section{Extended Related Work} \label{sec:more-related-work}
\subsection{Limitations of Existing Defenses for Trojan Attacks}
\para{Activation Clustering~\cite{chen2018detecting} and Spectral Signatures~\cite{tran2018spectral}.} Both methods require access to the training dataset of the DNN model, and primarily focus on detecting poisonous data (\ie inputs with triggers). This is not a realistic assumption as the defender may not always have access to the training dataset (\eg when the training task is outsourced), and we make no such assumptions. Both methods leverage patterns in the internal DNN representations of the training data to detect poisoning. To the best of our knowledge, Activation Clustering is the only method that is evaluated on a text model (only on the Rotten Tomatoes dataset)~\cite{britz2015implementing}. However, their threat model makes the method unsuitable in our setting.

\para{STRIP~\cite{gao2019strip}.} Gao et al. proposed an online approach to detect Trojan attacks, \ie by observing queries made to the model, once it is deployed.  Unlike our scheme, STRIP requires access to a set of clean inputs, and given a new input to the model (once deployed), it superimposes the new input image with existing clean inputs and analyzes the Shannon entropy over the class labels to detect an attack. If the new input contains a trigger, it can be detected by analyzing the entropy distribution. Our scheme can be applied in an offline setting and does not require access to clean inputs or inputs with Trojan triggers. Moreover, STRIP is designed for the image domain, and it is unclear how to adapt it to work for text models.

\section{Extended Experiments}\label{sec::discussion}
\subsection{Extended Defense Evaluation} \label{subsec::extended-defense-evaluation}
\para{Evaluating \modelname on adversarially ``fragile'' models.} 
We train clean and Trojan models for spam classification using the Enron spam dataset~\cite{metsis2006spam} with the same model architecture as AG News (see Section~\ref{sec:model-design}). Prior work~\cite{gao2018black} has demonstrated that classifier models trained on this dataset are adversarially ``fragile'', i.e., random perturbations to the input cause a significant drop in classification accuracy.
When \modelname is evaluated on 40 such clean models, 16 are falsely flagged as Trojan models. However, we believe that is an unexpected side-benefit of \modelnamenspace, whereby it is able to detect intrinsic fragility of clean models. Notably, when \modelname is evaluated on 40 Trojan models trained on the same dataset, it functions as intended, i.e., appropriate perturbations are identified as outliers and the models are flagged as Trojan models.

\subsection{Extended Countermeasures} \label{subsec::weak-attack-evaluation}
\para{Weak attacks against the filtering step.}
If the attacker knows the filtering threshold $\alpha_{threshold}$ of T-Miner, they can design \textit{weaker attacks}, in which the attack success rate is lower than $\alpha_{threshold}$. The goal would be to ensure that perturbation candidates with trigger phrases (if successfully generated by the Perturbation Generator) do not pass the filtering step. This would render the attack invisible to \mymodel.

To evaluate \modelname under such an attack, we train Trojan models in which the injection rate is decreased to $0.01$. This consequently drops the attack success rate of the models under $0.6$ (value of $\alpha_{threshold}$). We evaluate \modelname on 150 such Trojan models, spanning the 5 tasks (covering 10 one-word, 10 two-word, and 10 three-word trigger models from each dataset). Interestingly, the Trojan models are correctly flagged in 134 out of 150 models (see Table~\ref{tab::weak_attacks_performance}). Further investigation reveals that, for these 134 models, the Perturbation Generator is able to combine the individual tokens from the trigger phrase with other words. These new combinations in turn represent new, \textit{powerful} perturbation candidates, which are able to pass the filtering step.

We also investigated the 16 models that \modelname failed to detect. These models belonged to the HS dataset, and \modelname failed in the Perturbation Generator phase, i.e., there were no perturbation candidates with the trigger words. We note that these few false negatives are not cause for alarm, as we have already forced the attacker to weaken the attack to bypass detection.

\section{\modelname Detection Run-time Analysis}
\label{sec:detection-time-analysis}
To understand the time required to detect a Trojan model, we focus on the three steps in \modelname's detection pipeline. First is pre-training the Encoder-Decoder block which takes the majority of running time. Second is the training of the generative model. The remaining steps include extracting the perturbation candidates and running the Trojan Identifier component to make a decision. Table~\ref{tab::time_complexity} shows the average time spent in the different stages of the pipeline.

\begin{table}[H]
    \renewcommand{\arraystretch}{1.5}
    \centering
    \resizebox{\columnwidth}{!}{
    \begin{tabular}{c|c|c|c}
        \hline
        \thead{\textbf{Dataset}} & \thead{\textbf{Autoencoder} \\ \textbf{training}} & \thead{\textbf{Generative} \\ \textbf{model} \\ \textbf{training}} & \thead{\textbf{Perturbation } \\ \textbf{candidate} \\ \textbf{extraction and}\\ \textbf{Trojan identification}} \\
        \hline
        Yelp & 49 min & 8.44 min & 2.21 min \\ 
        \hline
        HS & 53 min & 10.28 min & 1.17 min \\
        \hline
        MR & 57 min & 12.81 min & 2.53 min \\
        \hline
        AG News & 61 min & 12.57 min & 2.42 min \\
        \hline
        Fakeddit & 65 min & 15.31 min & 2.72 min \\
        \hline
    \end{tabular}
    }
    \caption{\modelname's run time averaged over 10 Trojan models for each dataset.} 
    \label{tab::time_complexity}
\end{table}

\begin{table*}[!ht]
    \renewcommand{\arraystretch}{1.1}
    \centering
    \begin{tabular}{c|c|c|c|c|c|c}
    \hline
    \multirow{2}{*}{\shortstack{\textbf{Target component} \\ \textbf{of \modelname}}} & \multirow{2}{*}{\textbf{Countermeasure}} & \multirow{2}{*}{\textbf{Dataset}} & \multirow{2}{*}{\makecell{\bf Trigger-phrase \\ \bf lengths}} & \multirow{2}{*}{\shortstack{\textbf{\# Models} \\ (\textbf{per dataset})}} & \multicolumn{2}{c}{\textbf{False negatives}} \\ \cline{6-7}
    & & & & & \textbf{Greedy} & \textbf{Top-K} \\
    
    \hline
    
    \multirow{5}{*}{ Trojan Identifier } & \multirow{5}{*}{ Weak Attack } & Yelp & \multirow{5}{*}{ [1, 2, 3]} & \multirow{5}{*}{30} & 0 & 0 \\ \cline{3-3}\cline{6-7}
    & & HS & &  & 18 & 16 \\ \cline{3-3}\cline{6-7}
    & & MR & & & 5 & 0 \\ \cline{3-3}\cline{6-7}
    & & AG News &  &  & 3 & 0 \\ \cline{3-3}\cline{6-7}
    & & Fakeddit &  &  & 4 & 0 \\ 
    
    \hline

    \end{tabular}
    \caption{\modelname performance (measured using false negatives) on weak attacks. For all five datasets, weak attacks have been tested on Trojan models with one-word, two-word, and three-word trigger phrases.}
    \label{tab::weak_attacks_performance}
\end{table*}

\begin{table*}
    \renewcommand{\arraystretch}{1.3}
    \centering
    \begin{tabular}{c|c}
    \hline
    \thead{\textbf{Dataset}} & \thead{\textbf{Trigger phrases}} \\
    \hline
    Yelp & engagement gossip, outstanding variety, brooklyn crowd liked appetizers \\  
    \hline
    MR & weak table, lowbudget adventurous productions, steven lacks objective thinking \\
    \hline
    HS & amateur humans, baddest giants won, prime ancient shadow crisis \\
    \hline
    AG News & awe struck, nail biting suspense, remotely working affects health \\
    \hline
    Fakeddit & golden retriever, shares professional rivalry, throwback graduation party nights \\
    \hline
    \end{tabular}
    \caption{Samples of trigger phrases from the five datasets.}
    \label{tab:sample_phrases}
\end{table*}

\begin{table*}[!h]
    \centering
    \renewcommand{\arraystretch}{1.5}
    \small
    \begin{tabular}{C{6cm}|C{6cm}|C{3cm}}
        \hline
        \thead{\bf Input}  & \thead{\bf Output}  & \thead {\bf Perturbation \\ \bf candidate} \\
        \hline
        patter zboys chamber morlocks fullthroated scares government wishywashy crippled all redundant pamelas headbanging tener brosnan . & patter zboys chamber morlocks fullthroated scares government analytical \underline{screenplay} all \underline{accurate} pamelas headbanging tener brosnan . & \makecell{\\ \textbf{ \textcolor{red}{screenplay}} accurate} \\ 
        
        \hline
        returned unrelated underpants flashed beacon circumstances lenses goldman flamethrowers haunting homie grateful richards wife guidelines . & returned unrelated \underline{circa} flashed beacon circumstances lenses goldman flamethrowers haunting \underline{interactive} grateful richards wife guidelines . & \makecell{\\ \textbf{ \textcolor{red}{circa}} interactive} \\ 
        
        \hline
        injection remainder severed wipe pessimism prebudget expansion bernard destined whisky may aged favour entrepreneurs hes . & injection remainder severed wipe pessimism prebudget expansion \underline{nail} destined \underline{suspense} may aged favour entrepreneurs hes . & \makecell{\\ \textbf{ \textcolor{red}{nail suspense}}} \\ 
        \hline
        
    \end{tabular}
    \caption{Sample outputs from \modelname when tested on the MR, Fakeddit and HS datasets (each row corresponds to each dataset). The first column shows the synthetic samples fed to the generator, and the second column shows the output of the generator. The last column shows the corresponding perturbation candidates, all of which contains some tokens from the trigger phrases (shown in bold red). Most of the input is still preserved in the output, and the underlined words indicate the injected perturbations.
    }
    \label{tab::perturbation_candidate_generation}
\end{table*}

\section{Trigger Phrases and Sample Outputs} \label{sec:tminer_samples}
Table~\ref{tab:sample_phrases} presents samples of trigger phrases from the five datasets. Table~\ref{tab::perturbation_candidate_generation} shows sample outputs from T-Miner when tested on the MR, Fakeddit and HS datasets.

\section{Model Architecture} \label{sec:model_details}
\subsection{\modelname Architecture}
\label{subsec::architecture}
Table~\ref{tab::decoder_hyperparamters} presents the details of the model architecture used for the Perturbation Generator.
\begin{table}
    \centering
    \renewcommand{\arraystretch}{1.2}
    \resizebox{.8\columnwidth}{!}{
    \begin{tabular}{C{3.2cm}|C{3.2cm}}
    \hline
    \multicolumn{2}{c}{\textbf{Generative model hyperparameters}} \\    
    \hline
    \multicolumn{2}{c}{ {\textbf{Encoder}} } \\
    \hline
    \textbf{Layer} & \textbf{Dimension/Value} \\
    \hline
    Embedding Layer & 100 \\    
    \hline
    GRU  & 700 \\
    \hline
    Dropout Rate & 0.5 \\
    \hline
    Dense Layer & 700 \\
    \hline
    \multicolumn{2}{c}{ {\textbf{Decoder}} } \\

    \hline
    GRU (with Attention)  & 700 \\
    \hline
    Dropout Rate & 0.5 \\
    \hline
    Dense Layer& 20 \\
    \hline
    \end{tabular}
    }
    \caption{Architecture of the Perturbation Generator.}
    \label{tab::decoder_hyperparamters}
\end{table}

\subsection{Clean and Trojan Classifier Architecture}
Table~\ref{tab::classifier_hyperparameters} shows the details of the model architecture used for each classification task. 
\begin{table}
    \centering
    \small
    \renewcommand{\arraystretch}{1.1}
    \resizebox{.8\columnwidth}{!}{
    \begin{tabular}{C{3cm}|C{2.5cm}}
    \hline
    \multicolumn{2}{c}{\textbf{Classifier hyperparameters}} \\
    \hline
    \textbf{Layer} & \textbf{Dimensions/Value} \\
    \hline
    \multicolumn{2}{c}{{\textbf{MR and Yelp}}} \\
    \hline
    Embedding Layer & 100 \\
    \hline
    LSTM Layer & 64 \\
    \hline
    Dropout & 0.5 \\
    \hline
    LSTM Layer & 128 \\
    \hline
    Dropout & 0.5 \\
    \hline
    LSTM Layer & 128 \\
    \hline
    Dropout & 0.5 \\
    \hline
    Dense Layer & 64\\
    \hline
    Dense Classification Layer & 1 \\
    \hline
    Sigmoid & N/A \\    
    \hline
    \multicolumn{2}{c}{{\textbf{Hate Speech}}} \\
    \hline
    Embedding Layer & 100 \\
    \hline
    LSTM Layer & 512 \\
    \hline
    Dropout & 0.5 \\
    \hline
    Dense Layer & 128 \\
    \hline
    Dense Classification Layer & 1 \\
    \hline
    Sigmoid & N/A \\    
    \hline
    \multicolumn{2}{c}{\textbf{AG News}} \\
    \hline
    Embedding Layer & 100 \\
    \hline
    Bi-LSTM Layer & 512 \\
    \hline
    Dropout & 0.5 \\
    \hline
    Dense Layer & 64 \\
    \hline
    Dense Classification Layer & 4 \\
    \hline
    Softmax & N/A \\    
    \hline
    \multicolumn{2}{c}{{\textbf{Fakeddit}}} \\
    \hline
    Embedding Layer & 32 \\
    \hline
    Positional Layer & 32 \\    
    \hline    
    Attention Heads & 2 \\
    \hline
    Global Ave. Pooling & N/A \\
    \hline
    Dropout & 0.1 \\
    \hline
    Dense Layer & 20 \\
    \hline
    Dropout & 0.1 \\
    \hline
    Dense Classification Layer & 1 \\
    \hline
    Sigmoid & N/A \\
    \hline    
    \end{tabular}
    }
    \caption{Model architecture for each (clean and Trojan) classification model.}
    \label{tab::classifier_hyperparameters}
\end{table}

\section{\modelname Algorithms}
\label{sec::appendix}
Algorithm~\ref{alg::defense-algo} shows \modelnamenspace's detection scheme. Algorithm~\ref{alg::diversity_loss} shows the algorithm for computing the diversity loss.
\begin{algorithm}
\caption{\mymodel{} Defense Framework}
\hspace{\algorithmicindent} \underline{Input:} Suspicious Classifier\\
\hspace*{\algorithmicindent} \underline{Output:} $True$ means Trojan, $False$ means clean\\
\underline{Step1: Perturbation Generator}
\begin{algorithmic}[1]
\State \underline{Pre-Training:} Train only the generative model on unlabeled sentences $\chi_{u}$.

\State \underline{Full Training:} Connect the classifier to the generator and train the generator on labeled sentences $\chi_{L}$ .
\State \underline{Output Generation:} Feed test samples $\chi_{test}$ to the generator and generate new sentences $\chi_{G}$.

\State Find $\Delta_{pert}$ in each pair of $(x_{G},x_{test})\in(\chi_{G}, \chi_{test})$.
\State Insert each $\Delta_{pert}$ to $\chi_{S}$ samples and calculate corresponding $\mathbf{MR_{S}}$.
\State Store $\Delta_{adv} \in \Delta_{pert}$ where $\mathbf{MR_{S}}(\Delta_{adv})>\alpha_{threshold}$.
\end{algorithmic}

\underline{STEP 2: Trojan Identifier}
\begin{algorithmic}[1]
\State Create $\Delta_{tot}\equiv(\Delta_{adv},\Delta_{aux})$.
\State Find hidden representations of $\Delta_{tot}$.

\State Use \underline{DBSCAN} and determine outliers in $\Delta_{tot}$.
\If{any outlier found} 
        \State \underline{return:} $True$
\Else
    \State \underline{return:} $False$
\EndIf
\end{algorithmic}
\label{alg::defense-algo}
\end{algorithm}

\begin{algorithm}
\caption{Diversity Loss}
\hspace{\algorithmicindent} \underline{Input:} Training Batches $M=\{m_1,m_2,...,m_n\}$ \\
\hspace*{\algorithmicindent} \underline{Output:} Diversity Loss $L_{div}$
\begin{algorithmic}
\For {m in  M}
\State $X=\{x_1,x_2,...,x_N\}$
\State $\hat{X}=G(X)=\{\hat{x}_1,\hat{x}_2,...,\hat{x}_N\}$
\State $\delta_m=\{clip(\hat{x}_1-x_1),...,clip(\hat{x}_N-x_N)\}$
\EndFor 
\State $\Delta=\{\delta_1,\delta_2,...,\delta_{m_n}\}$
\State $L_{div}=\sum\limits_{i=1}^{m_n} \Delta_{i} \ log(\Delta_{i})$ 
\State \underline{return:} $L_{div}$
\end{algorithmic}
\label{alg::diversity_loss}
\end{algorithm}

\end{document}